\documentclass[draftclsnofoot, 12pt,onecolumn,oneside,compress]{IEEEtran}

\ifCLASSINFOpdf
\else
   \usepackage[dvips]{graphicx}
\fi
\usepackage{url}

\usepackage{graphicx,balance}
\usepackage{bm}
\usepackage{amssymb}
\usepackage{amsmath}
\usepackage{textcomp}
\usepackage{color}
\usepackage{epstopdf}
\usepackage{subfigure}
\usepackage{booktabs}
\usepackage{enumitem}
\usepackage{algorithm}
\usepackage{algorithmic}
\usepackage{bbm}
\usepackage{makecell}
\usepackage{multirow}

\begin{document}

\title{Multi-User Matching and Resource Allocation in Vision Aided Communications}

\author{Weihua Xu, Feifei Gao, Yong Zhang, Chengkang Pan, and Guangyi Liu
\thanks{W. Xu and F. Gao are with Institute for Artificial Intelligence, Tsinghua
University (THUAI), Beijing National Research Center for Information
Science and Technology (BNRist), Department of Automation, Tsinghua
University, Beijing, P.R. China, 100084 (email: xwh19@mails.tsinghua.edu.cn,
feifeigao@ieee.org).}
\thanks{Y. Zhang is with the Beijing Key Laboratory of Multimedia and Intelligent Software Technology, Beijing Institute of Artificial Intelligence, Faculty of Information Technology, Beijing University of Technology, Beijing, P.R. China, 100084 (e-mail: zhangyong2010@bjut.edu.cn).}
\thanks{C. Pan and G. Liu are with the China Mobile Research Institute, Beijing 100053, P.R. China (e-mail: panchengkang@chinamobile.com; liuguangyi@chinamobile.com).}
}

\maketitle
\vspace{-5mm}
\begin{abstract}
Visual perception is an effective way to obtain the spatial characteristics of wireless channels and to reduce the overhead for communications system. A critical problem for the visual assistance is that the communications system needs to match the radio signal with the visual information of the corresponding user, i.e., to identify the visual user that corresponds to the target radio signal from all the environmental objects. In this paper, we propose a user matching method for environment with a variable number of objects. Specifically, we apply 3D detection to extract all the environmental objects from the images taken by multiple cameras. Then, we design a deep neural network (DNN) to estimate the location distribution of users by the images and beam pairs at multiple moments, and thereby identify the users from all the extracted environmental objects. Moreover, we present a resource allocation method based on the taken images to reduce the time and spectrum overhead compared to traditional resource allocation methods. Simulation results show that the proposed user matching method outperforms the existing methods, and the proposed resource allocation method can achieve $92\%$ transmission rate of the traditional resource allocation method but with the time and spectrum overhead significantly reduced.
\end{abstract}


\IEEEpeerreviewmaketitle

\newpage

\section{Introduction}
The environmental information, such as object sizes, locations and orientations, can effectively indicate the wireless propagation characteristics \cite{YCui}-\cite{FLiu}. With the progressive development of intelligent business, such as intelligent transportation and autonomous driving, abundant types of sensors, including Radar, LIDAR and Depth/RGB cameras, will be mounted on the communications equipment or the infrastructures \cite{Heath}-\cite{AAli}. These sensors can be explored to assist many communications tasks, such as beam alignment \cite{AKlautau}-\cite{WXu1}, channel covariance matrix estimation \cite{WXu2}, as well as the prediction for blockage, base station (BS) handover, as well as received power \cite{YOguma}-\cite{TNishio}. Among them, the visual perception by cameras has drawn much attention recently, due to its universality, low cost, and high resolution compared with the Radar and LIDAR.

Naturally, user matching, i.e., identifying the visual user from all the environmental objects, is a critical problem for the visual aided communications. However, the authors of \cite{YOguma}-\cite{TNishio} assume the user is static. The authors of \cite{MAlrabeiah} design a vision aided service identification method, and the authors of \cite{GCharan2} propose a vision based beam selection method. Nevertheless, in \cite{MAlrabeiah}-\cite{GCharan2}, only the user can move, and all the surrounding objects are static. The authors of \cite{YTian}-\cite{GCharan1} focus on the traffic scenario with multiple dynamic vehicles, and propose a vision based beam tracking and blockage/handover prediction method. However, in \cite{YTian}-\cite{GCharan1}, the user matching task is performed together with the communication task by a single DNN, which requires the user characteristic information, such as the beam sequence, for user matching. Note that, acquiring the user characteristic
information will lead to ineluctable communications overhead. Moreover, the accurate characteristic information can not always be obtained, which fails the algorithms designed in \cite{YTian}-\cite{GCharan1}.

To achieve better robustness and lower cost, it is more reasonable to separately performs the user matching task and the communications task. Once the user has been matched to its visual image, the latter can be tracked continuously by the state-of-the-art object tracking techniques \cite{AYilmaz}-\cite{BDeori} without any communications overhead. To handle the user matching task, the authors in \cite{VMPinho} design a classification DNN to distinguish user from all the environmental objects by the channel of the user and the detected object bounding boxes (BBoxes) in the image. Nevertheless, the algorithm of \cite{VMPinho} can only handle two environmental objects, whereas the object number is random in realistic environment, and the channel is hard to obtain especially for the system with large scale antenna arrays.

In this paper, we propose a vision aided communication scheme that includes a user matching method and a resource allocation method. The proposed user matching method can adapt to the communications environment with a variable number of dynamic objects, and the proposed resource allocation method can realize power allocation and user scheduling by the vision information. The frame diagram of the proposed vision aided communication scheme as well as the frame diagram of the compared methods is shown in Fig.~1. The main contributions are as follows:
\begin{figure}[t]
\centering
\includegraphics[width=0.8\textwidth]{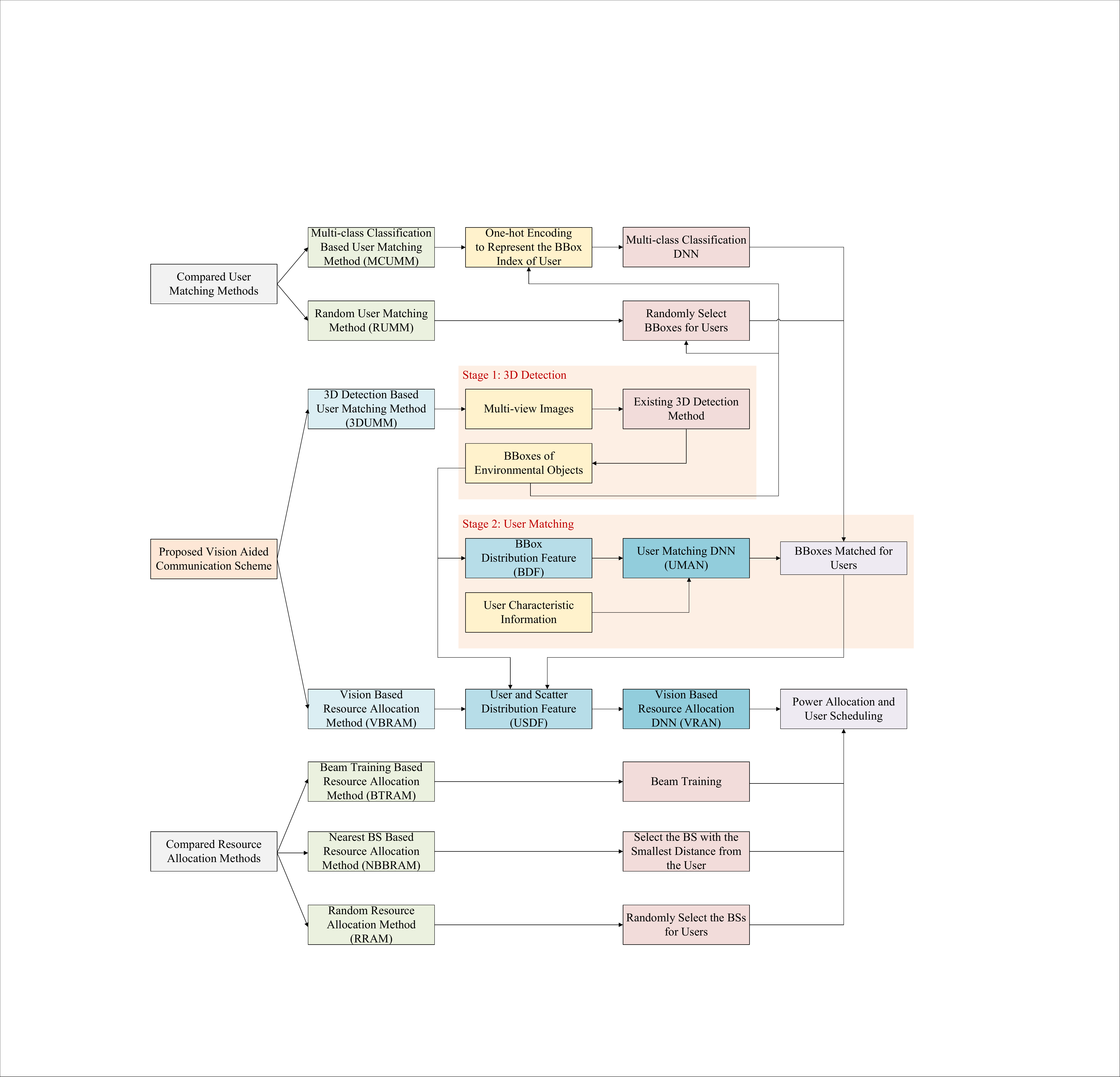}
\caption{The frame diagram of the proposed vision aided communication scheme as well as the frame diagram of the compared methods. The scheme includes a user matching method and a resource allocation method.}
\end{figure}
\begin{itemize}
    \item[1)]
    3D detection based user matching method (3DUMM): We firstly use the 3D detection method to obtain the location/size/orientation information of all environmental objects from the multi-view images. Then, we utilize the the spatial distribution information of all the detected objects to generate a BBox distribution feature (BDF). We design a user matching DNN (UMAN) to estimate the location distribution of targeted user by the BDF and the optimal beam pair indices. Thus, we identify the visual user by searching the environmental object with the minimum distance from the user location.
    \item[2)]
    Vision based resource allocation method (VBRAM): For the scenario with multiple BSs and multiple users, we use the spatial distribution of the users and the surrounding scattering objects to generate a user and scatter distribution feature (USDF). Then, we design a vision based resource allocation DNN (VRAN) to predict both the optimal BS that each user should connect and the optimal transmission power by USDF.
    \item[3)]
    Simulation Dataset Generation: We construct a 3D environment model with multiple dynamic vehicles and utilize ray tracing for channel generation. With the simulated data, the proposed 3DUMM shows better matching accuracy than the existing user matching method, and the proposed VBRAM could achieve comparable performance as the conventional resource allocation method.
\end{itemize}

This paper is organized as follows. Section II introduces the signal model. Section III describes the 3D detection based user matching method, and Section IV proposes the vision based resource allocation method. Section V presents the simulation setup, performance metric, numerical results, and detailed discussions. Finally, the conclusions are given in Section VI.

Notation: $\bm{A}$ is a matrix or tensor; $\bm{\mathcal{A}}$ is a set; $\bm{a}$ is a vector; $a$ is a scalar; $\bm{A}[i,j]$ is the element of the $i$th row and the $j$th column in $\bm{A}$; $\bm{A}[i,:]$ and $\bm{A}[:,j]$ are the $i$th row and the $j$th column of $\bm{A}$ respectively; $\mathcal{N}(\bm{m}_{\mathrm{g}}, \bm{R}_{\mathrm{g}})/\mathcal{CN}(\bm{m}_{\mathrm{g}}, \bm{R}_{\mathrm{g}})$ is the real/complex Gaussian
random distribution with mean $\bm{m}_{\mathrm{g}}$ and covariance $\bm{R}_{\mathrm{g}}$; $\mathrm{Card}(\bm{\mathcal{A}})$ is the cardinality of the set $\bm{\mathcal{A}}$. $\lfloor a \rfloor$ is the maximum integer not exceeding $a$. $\lceil a \rceil$ is the minimum integer not less than $a$. $\bm{\mathcal{A}}\backslash \bm{\mathcal{A}}_{\mathrm{g}}$ is the set $\bm{\mathcal{A}}$ in which all the elements of $\bm{\mathcal{A}}_{\mathrm{g}}$ are removed. The key notations are summarized in Table~I.

\begin{table}[t]
\scriptsize
\centering
\caption{key notations in the paper}
\begin{tabular}{|c|c|c|c|}
\hline
\textbf{Notation}& \textbf{Description}& \textbf{Notation}& \textbf{Description}\\
\hline
$B$ and $U$ & Numbers of the BSs and users& $L_{\mathrm{D}}$ and $W_{\mathrm{D}}$ & Length and width of the grid used for $\bm{D}_k$\\
\hline
$N_{\mathrm{B}}$ and $N_{\mathrm{U}}$ & Antenna numbers of the BS and user& $N_{\mathrm{DX}}$ and $N_{\mathrm{DY}}$& \makecell{Number of columns and rows\\of the grids used for $\bm{D}_k$}\\
\hline
$b_u$ & Index of the BS that serves the $u$th user&  $\bm{\mathcal{X}}_{\mathrm{E},k}^{n_{\mathrm{DX}},n_{\mathrm{DY}}}$ &  \makecell{Set formed by counting which BBoxes in\\ $\bm{\mathcal{X}}_{\mathrm{E},k}$ are inside the $(n_{\mathrm{DX}},n_{\mathrm{DY}})$th grid used for $\bm{D}_k$}\\
\hline
$\bm{H}_{b,u}$ & \makecell{Downlink channel matrix\\between the $b$th BS
and the $u$th user}& $M$& \makecell{Length of BBox distribution feature sequence\\and beam pair index sequence for 3DUMM}\\
\hline
$P_u$ & \makecell{Transmission power of the $b_u$th BS}& $\bm{F}_k$ & Keypoint heatmap for 3DUMM\\
\hline
$(\bm{\mathrm{f}}^{\mathrm{B},\mathrm{opt}}_{b,u},\bm{\mathrm{f}}^{\mathrm{U},\mathrm{opt}}_{b,u})$ & Optimal beam pair for the channel $\bm{H}_{b,u}$& $L_{\mathrm{F}}$ and $W_{\mathrm{F}}$ & Length and width of the heatmap grid used for $\bm{F}_k$\\
\hline
$N_{\mathrm{CB}}$ and $N_{\mathrm{CU}}$ &Codebook sizes of transmit and receive beam& $N_{\mathrm{FX}}$ and $N_{\mathrm{FY}}$& \makecell{Number of columns and rows\\of the heatmap grids used for $\bm{F}_k$}\\
\hline
$\mathrm{Box}_{b,v}$ &\makecell{3D BBox of\\the $v$th vehicle in the $b$th image}&  $\mathrm{Box}_k$ & BBox in $\bm{\mathcal{X}}_{\mathrm{E},k}$ of the target user for 3DUMM\\
\hline
$(x^{\mathrm{G}}_{b,v}, y^{\mathrm{G}}_{b,v}, z^{\mathrm{G}}_{b,v})$ &\makecell{Center location coordinates of $\mathrm{Box}_{b,v}$\\under the global coordinate system} &$(n_{\mathrm{X},k},n_{\mathrm{Y},k})$  &\makecell{Index of the heatmap grid that contains $\mathrm{Box}_k$}\\
\hline
$l_{b,v}$,$w_{b,v}$ and $h_{b,v}$ & Length, width and height of $\mathrm{Box}_{b,v}$ & $\bm{Z}_k$ & User and scatter distribution feature for VBRAM\\
\hline
$V_{\mathrm{O}}$ &\makecell{Overlap volume of two 3D\\BBoxes $\mathrm{Box}_{b,v}$ and $\mathrm{Box}_{b^{'},v^{'}}$} & $L_{\mathrm{R}}$ and $W_{\mathrm{R}}$ & Length and width of the grid used for $\bm{Z}_k$\\
\hline
$T_{\mathrm{b}}$ and $T_{\mathrm{f}}$ & \makecell{Beam coherent time and\\Shooting interval of the cameras} & $N_{\mathrm{RX}}$ and $N_{\mathrm{RY}}$& \makecell{Number of columns and rows\\of the grids used for $\bm{Z}_k$}\\
\hline
$\bm{\mathcal{X}}_{\mathrm{E},k}$ & \makecell{De-redundancy BBox set\\obtained at the moment $kT_{\mathrm{b}}$}& $\mathrm{Box}_{k}^{u}$& BBox of the $u$th user in $\bm{\mathcal{X}}_{\mathrm{E},k}$ for VBRAM\\
\hline
$I_{u,k}$ &\makecell{Index of the optimal beam pair\\at the moment $kT_{\mathrm{b}}$ for the $u$th user}& $\bm{\mathcal{X}}_{\mathrm{RE},k}^{n_{\mathrm{RX}},n_{\mathrm{RY}}}$ &  \makecell{Set formed by counting which BBoxes in\\ $\bm{\mathcal{X}}_{\mathrm{E},k}$ are inside the $(n_{\mathrm{FX}},n_{\mathrm{FY}})$th grid used for $\bm{Z}_k$}\\
\hline
$\bm{D}_k$ &BBox distribution feature for 3DUMM& $\bm{O}^{\mathrm{B}}_{k}$ and $\bm{O}^{\mathrm{P}}_{k}$& \makecell{Matrices to represent the optimal allocation\\of BSs and transmission power}\\
\hline
\end{tabular}
\end{table}

\section{Signal Model}
We consider a downlink mmWave communications system with $B$ BSs and $U$ users. Each BS and each user are equipped with a uniform linear array (ULA) of $N_\mathrm{B}$ and $N_\mathrm{U}$ antennas respectively. For simplicity, we assume both BS and MS have one mmWave radio frequency chain and $U\leq B$. Hence, each BS is assumed to serve at most a single user on a single frequency band. Nevertheless, the proposed scheme can be straightforwardly extended to the communications scenario where a single BS serves multiple users with frequency-division multiplexing or the multi-carrier systems. Denoting the index of the BS that serves the $u$th user as $b_{u}$, where $b_u\in\bm{\mathcal{B}}=\{1,2,\cdots,B\}$, the downlink signal received at the $u$th user can be expressed as
\begin{equation}
y_{u}=(\bm{\mathrm{f}}^{\mathrm{U}}_{u})^{\mathrm{H}}\bm{H}_{b_{u},u}\bm{\mathrm{f}}^{\mathrm{B}}_{b_u}s_{u}+\sum_{u^{'}\in \bm{\mathcal{U}}\backslash\{u\}}(\bm{\mathrm{f}}^{\mathrm{U}}_{u})^{\mathrm{H}}\bm{H}_{b_{u^{'}},u}\bm{\mathrm{f}}^{\mathrm{B}}_{b_{u^{'}}}s_{u^{'}}+n_{u},
\end{equation}
where $\bm{\mathcal{U}}=\{1,2,\cdots,U\}$, $\bm{H}_{b,u}\in \mathbb{C}^{N_{\mathrm{U}}\times N_{\mathrm{B}}}$ is the downlink channel matrix between the $b$th BS and the $u$th user, $s_{u}\in \mathbb{C}$ is the transmit signal of the $b_u$th BS, $\bm{\mathrm{f}}^{\mathrm{B}}_{b}\in \mathbb{C}^{N_\mathrm{B}\times 1}$ is the transmit beamforming vector of the $b$th BS, $\bm{\mathrm{f}}^{\mathrm{U}}_{u}\in \mathbb{C}^{N_{\mathrm{U}}\times 1}$ is the receive beamforming vector of the $u$th user, and $n_{u} \in \mathcal{CN}(0,\sigma^2)$ is the Gaussian noise, $u=1,2,\cdots,U$, $b=1,2,\cdots,B$. The transmit signal $s_{u}$ has the power $\mathrm{E}\{|s_{u}|^2\}=P_{u}$.

We adopt the codebook based analog beamforming for the communications. The optimal beam pair between the users and the corresponding serving BSs, i.e., the $b_1$th, $b_2$th, $\cdots$, $b_U$th BS, can be selected by the beam training\footnote{The algorithm for beam alignment is independent to the proposed 3DUMM and can be arbitrary.}. Specifically, the $b_u$th BS can select the optimal beam pair $(\bm{\mathrm{f}}^{\mathrm{B},\mathrm{opt}}_{b_u,u},\bm{\mathrm{f}}^{\mathrm{U},\mathrm{opt}}_{b_u,u})$ from the transmit beam codebook $\bm{\mathcal{F}}_{\mathrm{B}}=\{\bm{\mathrm{f}}^{\mathrm{B},1}, \bm{\mathrm{f}}^{\mathrm{B},2}, \cdots, \bm{\mathrm{f}}^{\mathrm{B},N_{\mathrm{CB}}}\}$ and the receive beam codebook $\bm{\mathcal{F}}_{\mathrm{U}}=\{\bm{\mathrm{f}}^{\mathrm{U},1}, \bm{\mathrm{f}}^{\mathrm{U},2}, \cdots, \bm{\mathrm{f}}^{\mathrm{U},N_{\mathrm{CU}}}\}$ by maximizing the receive signal-to-noise ratio (SNR), i.e.,
\begin{equation}
(\bm{\mathrm{f}}^{\mathrm{B},\mathrm{opt}}_{b,u},\bm{\mathrm{f}}^{\mathrm{U},\mathrm{opt}}_{b,u})=\mathop{\arg\max}_{\bm{\mathrm{f}}^{\mathrm{B}}\in \bm{\mathcal{F}}_{\mathrm{B}},\bm{\mathrm{f}}^{\mathrm{U}}\in \bm{\mathcal{F}}_{\mathrm{U}}} \frac{1}{\sigma^2}|(\bm{\mathrm{f}}^{\mathrm{U}})^{\mathrm{H}}\bm{H}_{b,u}\bm{\mathrm{f}}^{\mathrm{B}}|^2.
\end{equation}

\section{3D Detection Based User Matching}
\begin{figure}[t]
\centering
\includegraphics[width=0.6\textwidth]{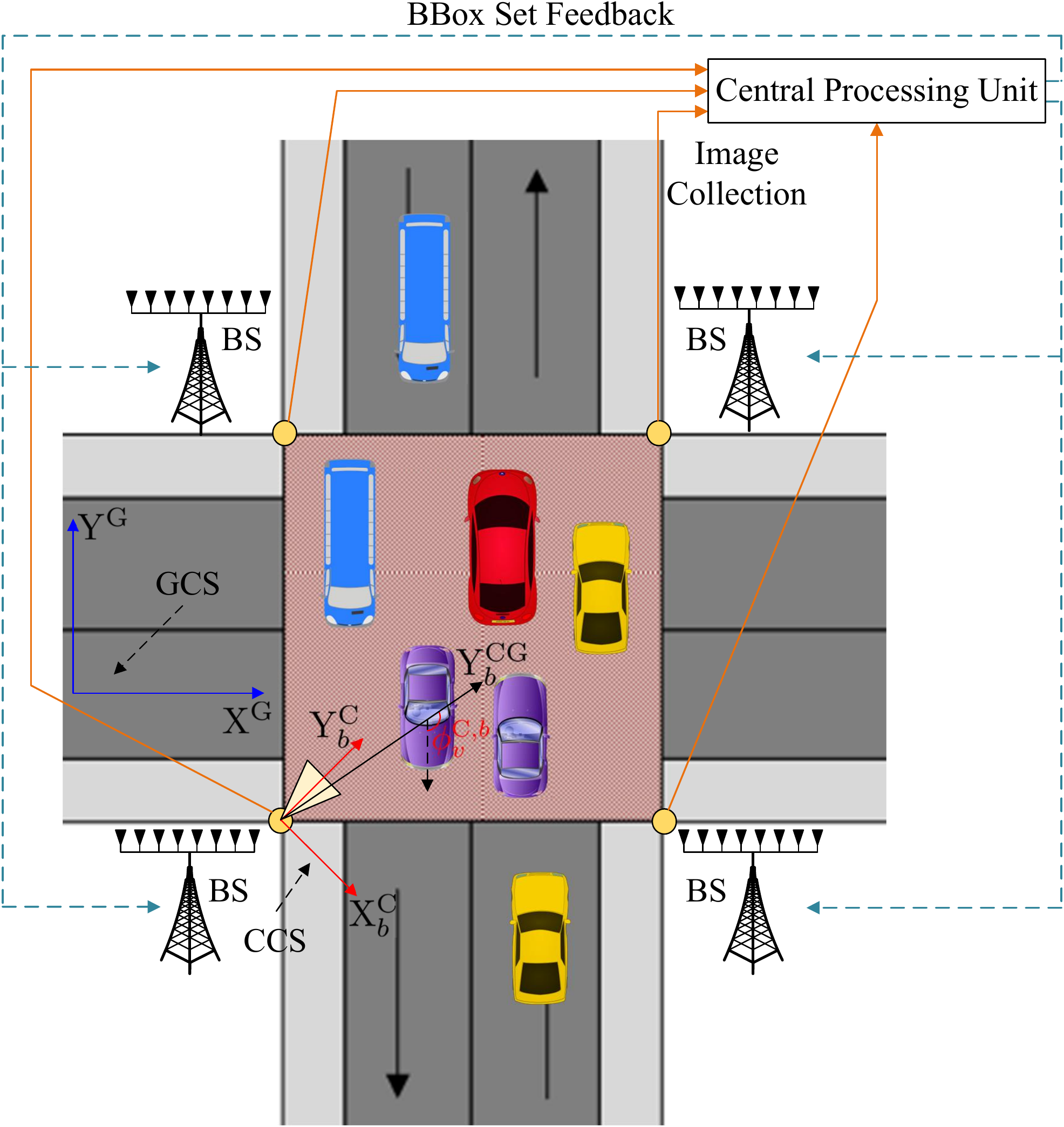}
\caption{The concerned communications scenario for 3DUMM. The communications scenario includes two traffic lanes and a random number of vehicles. BSs equipped with the cameras are deployed at the road-side to serve multiple vehicles (users) in the lanes.}
\end{figure}

For the vision aided communication system, the radio signal of each user needs to be matched with the corresponding user object in the visual images. Thus, to support the user matching, the environment objects that are probable user objects should be extracted from the images by the object detection techniques, including 2D detection \cite{YOLO} and 3D detection \cite{Zliu}-\cite{mv3d}. Compared with 2D detection, 3D detection can obtain the 3D locations, 3D sizes, and orientations of objects, and thereby provide more abundant spatial distribution information of environment objects for the communication system. Therefore, we propose to adopt the 3D detection to extract the 3D BBoxes of the environmental objects in the images and match the user radio signal with the corresponding user 3D BBox.

Fig.~2 shows the concerned communications scenario, in which multiple vehicles run in two traffic lanes, among which $U$ vehicles are communications users. Meanwhile, $B$ BSs are deployed at the road-side to serve these $U$ users, and BSs are all equipped with monocular RGB cameras to monitor the vehicles. All the cameras are connected to a central processing unit that gathers the images taken by all the cameras. Thus, the central processing unit can obtain the visual information of all the vehicles in the crossroad. We utilize the 3D object detection technique to accurately detect the location, the size, and the orientation of the vehicles.

\begin{figure}[t]
	\begin{minipage}[t]{0.5\linewidth}
		\centering
			\includegraphics[width=50mm]{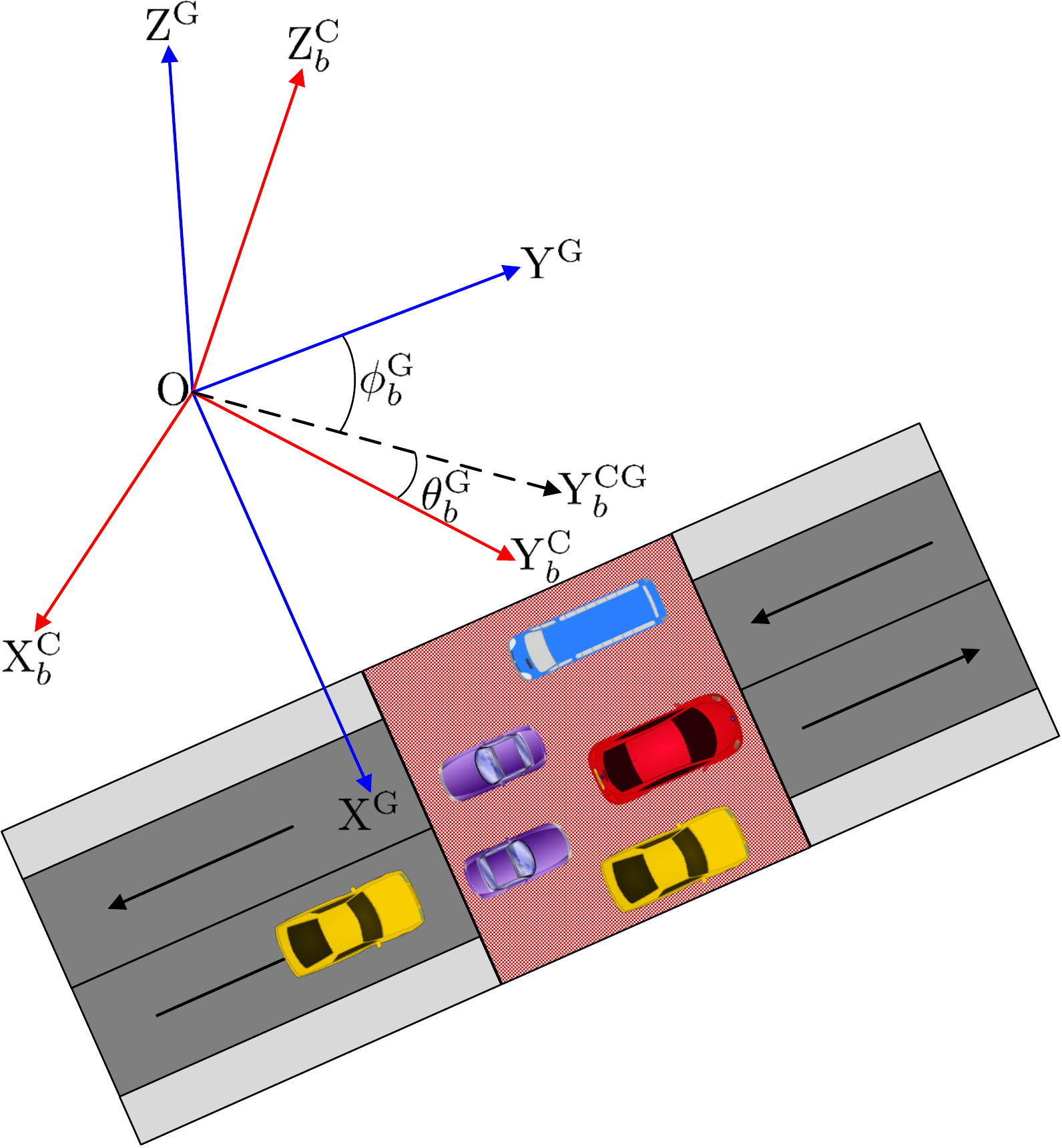}
		\caption{The coordinate relation between $b$th CCS and GCS. $\mathrm{Y}^{\mathrm{CG}}_{b}$ axis: the projection of the $\mathrm{Y}^{\mathrm{C}}_{b}$ axis on the $\mathrm{X}^{\mathrm{G}}$-$\mathrm{Y}^{\mathrm{G}}$ plane; $\theta^{\mathrm{G}}_{b}$: the elevation angle of the $b$th camera; $\phi^{\mathrm{G}}_{b}$: the azimuth angle of the $b$th camera.}
	\end{minipage}
	\hspace{1ex}
	\begin{minipage}[t]{0.5\linewidth}
		\centering
	\includegraphics[width=83mm]{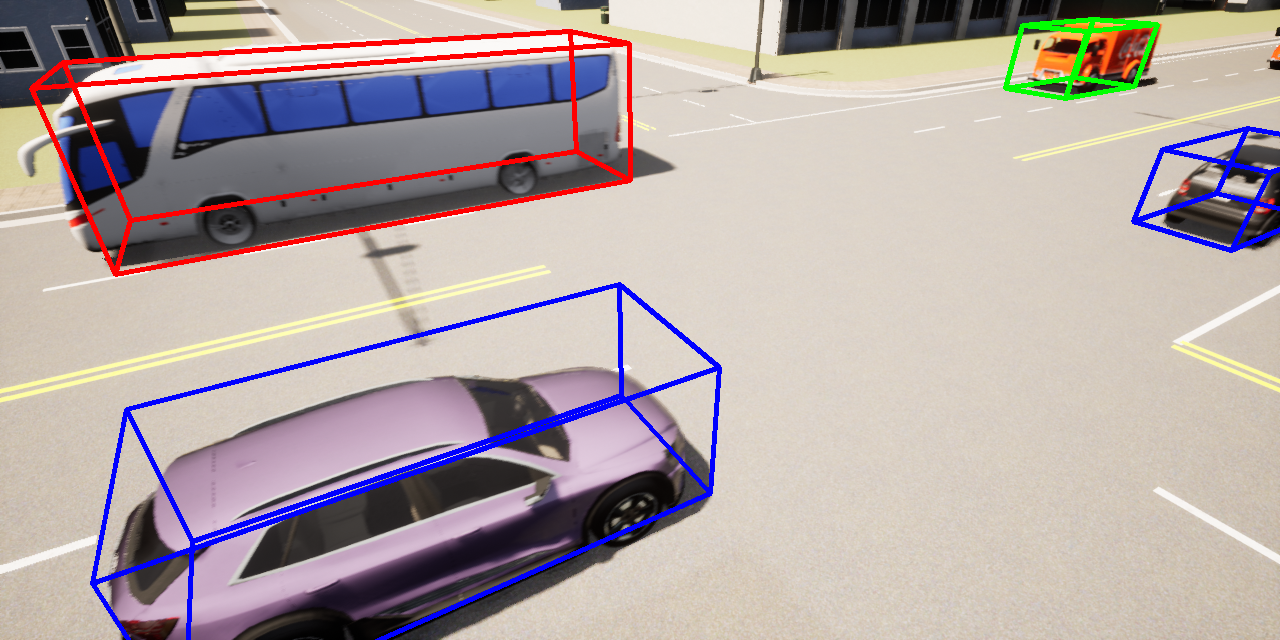}
		\caption{The vehicles' 3D BBoxes obtained by 3D detection method for one camera.}
	\end{minipage}
\end{figure}

For the camera in the $b$th BS, $b=1,2,\cdots,B$, we define the camera coordinate system (CCS) with $\mathrm{X}^{\mathrm{C}}_{b}$-axis, $\mathrm{Y}^{\mathrm{C}}_{b}$-axis, and $\mathrm{Z}^{\mathrm{C}}_{b}$-axis, where the origin and the $\mathrm{Y}^{\mathrm{C}}_{b}$-axis are the $b$th camera's location and the optic axis, respectively. Moreover, we define a global coordinate system (GCS) with $\mathrm{X}^{\mathrm{G}}$-axis, $\mathrm{Y}^{\mathrm{G}}$-axis, and $\mathrm{Z}^{\mathrm{G}}$-axis, where the $\mathrm{X}^{\mathrm{G}}$-axis and the $\mathrm{Y}^{\mathrm{G}}$-axis are parallel with the horizontal and the vertical edge of the crossroad respectively, while the origin can be selected arbitrarily on the plane of crossroad. We denote the projection of the $\mathrm{Y}^{\mathrm{C}}_{b}$ axis on the $\mathrm{X}^{\mathrm{G}}$-$\mathrm{Y}^{\mathrm{G}}$ plane as $\mathrm{Y}^{\mathrm{CG}}_{b}$-axis, as shown in Fig.~2 and Fig.~3. For each vehicle, we define the angle between the vehicle orientation and the $\mathrm{Y}^{\mathrm{CG}}_{b}$-axis as the azimuth of the vehicle in the corresponding CCS. Similarly, we define the angle between the vehicle orientation and the $\mathrm{Y}^{\mathrm{G}}$-axis as the azimuth of the vehicle in GCS. Note that, the superscript is used to distinguish which coordinate system is used for a coordinate value or an azimuth. For example, $(x^{\mathrm{C}}_b,y^{\mathrm{C}}_b,z^{\mathrm{C}}_b)$ and $(x^{\mathrm{G}},y^{\mathrm{G}},z^{\mathrm{G}})$ denote the coordinate values under the $b$th CCS and the GCS respectively, while $\phi^{\mathrm{C}}_b$ and $\phi^{\mathrm{G}}$ denote the azimuth in the $b$th CCS and the GCS respectively.

The proposed 3DUMM contains two stages: 3D detection and user matching. In the following subsections, we detail these stages.

\subsection{3D Detection Stage}
We here utilize the 3D detection method to obtain the 3D BBox of each vehicle in every image, as shown in Fig.~4. Thus, the task of user matching can be treated as matching the radio signal of each BS with the 3D BBox of the corresponding user. Specifically, we denote the number of vehicles in the $b$th image as $ N_{\mathrm{V},b}$ and denote the $v$th vehicle in the $b$th image as the $(b,v)$th vehicle. One can estimate the length $l_{b,v}$, the width $w_{b,v}$, the height $h_{b,v}$, the center location $(x^{\mathrm{C}}_{b,v},y^{\mathrm{C}}_{b,v},z^{\mathrm{C}}_{b,v})$, and the azimuth $\phi^{\mathrm{C}}_{b,v}$ of the $(b,v)$th vehicle by the 3D detection method, $v=1,2,\cdots,N_{\mathrm{V},b}$. Then, we obtain the location $(x_{b}^{\mathrm{G}},y_{b}^{\mathrm{G}},z_{b}^{\mathrm{G}})$, the elevation angle $\theta^{\mathrm{G}}_{b}$, and the azimuth angle $\phi^{\mathrm{G}}_{b}$ of the $b$th camera, $b\in \bm{\mathcal{B}}$, as the coordinate relation between the $b$th CCS and the GCS shown in Fig.~3. Thus, we can obtain the location $(x^{\mathrm{G}}_{b,v}, y^{\mathrm{G}}_{b,v}, z^{\mathrm{G}}_{b,v})$ and the azimuth $\phi^{\mathrm{G}}_{b,v}$ of the $(b,v)$th vehicle by the coordinate transformation as
\begin{equation}
\begin{aligned}
\left[
\begin{matrix}
  x^{\mathrm{G}}_{b,v} \\
  y^{\mathrm{G}}_{b,v} \\
  z^{\mathrm{G}}_{b,v}
\end{matrix}\right]&=\left[\begin{matrix}
\cos(\phi^{\mathrm{G}}_{b}) & \sin(\phi^{\mathrm{G}}_{b}) & 0\\
-\sin(\phi^{\mathrm{G}}_{b}) & \cos(\phi^{\mathrm{G}}_{b}) & 0\\
0 & 0 & 1
\end{matrix}\right]\left[\begin{matrix}
1 & 0 & 0 \\
0 & \cos(\theta^{\mathrm{G}}_{b}) & \sin(\theta^{\mathrm{G}}_{b}) \\
0 & -\sin(\theta^{\mathrm{G}}_{b}) & \cos(\theta^{\mathrm{G}}_{b})
\end{matrix}\right]
\left[
\begin{matrix}
  x^{\mathrm{C}}_{b,v} \\
  y^{\mathrm{C}}_{b,v} \\
  z^{\mathrm{C}}_{b,v}
\end{matrix}\right]+
\left[
\begin{matrix}
  x_{b}^{\mathrm{G}} \\
  y_{b}^{\mathrm{G}} \\
  z_{b}^{\mathrm{G}}
\end{matrix}\right],
\\
\phi^{\mathrm{G}}_{b,v}&=\phi^{\mathrm{G}}_{b}+\phi^{\mathrm{C}}_{b,v}, v=1,2,\cdots,N_{\mathrm{V},b}, b=1,2,\cdots,B.
\end{aligned}
\end{equation}

\begin{figure}[t]
	\begin{minipage}[t]{0.5\linewidth}
		\centering
			\includegraphics[width=83mm]{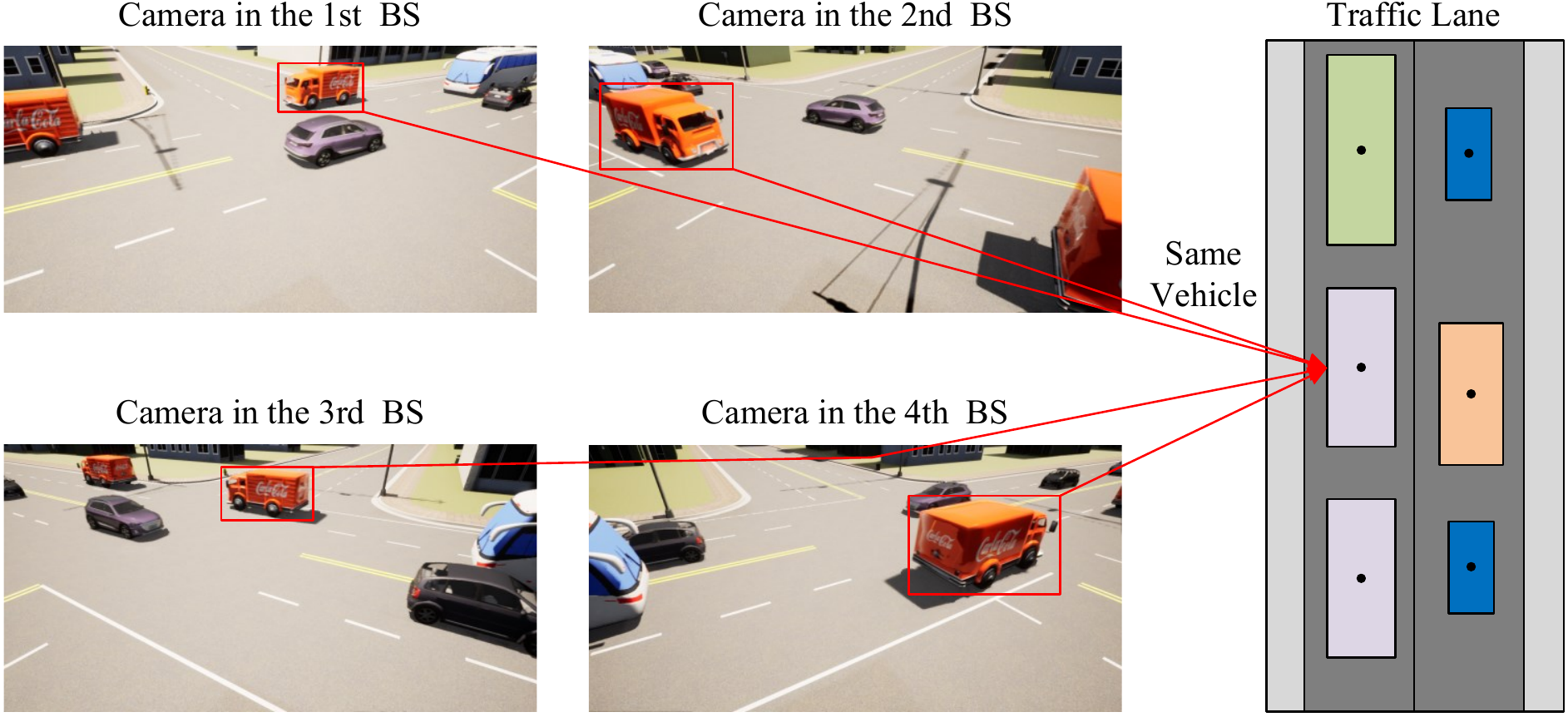}
		\caption{The overlap of FOV of different cameras, where the same van appears in different images taken by the different cameras.}
	\end{minipage}
	\hspace{1ex}
	\begin{minipage}[t]{0.5\linewidth}
		\centering
	\includegraphics[width=83mm]{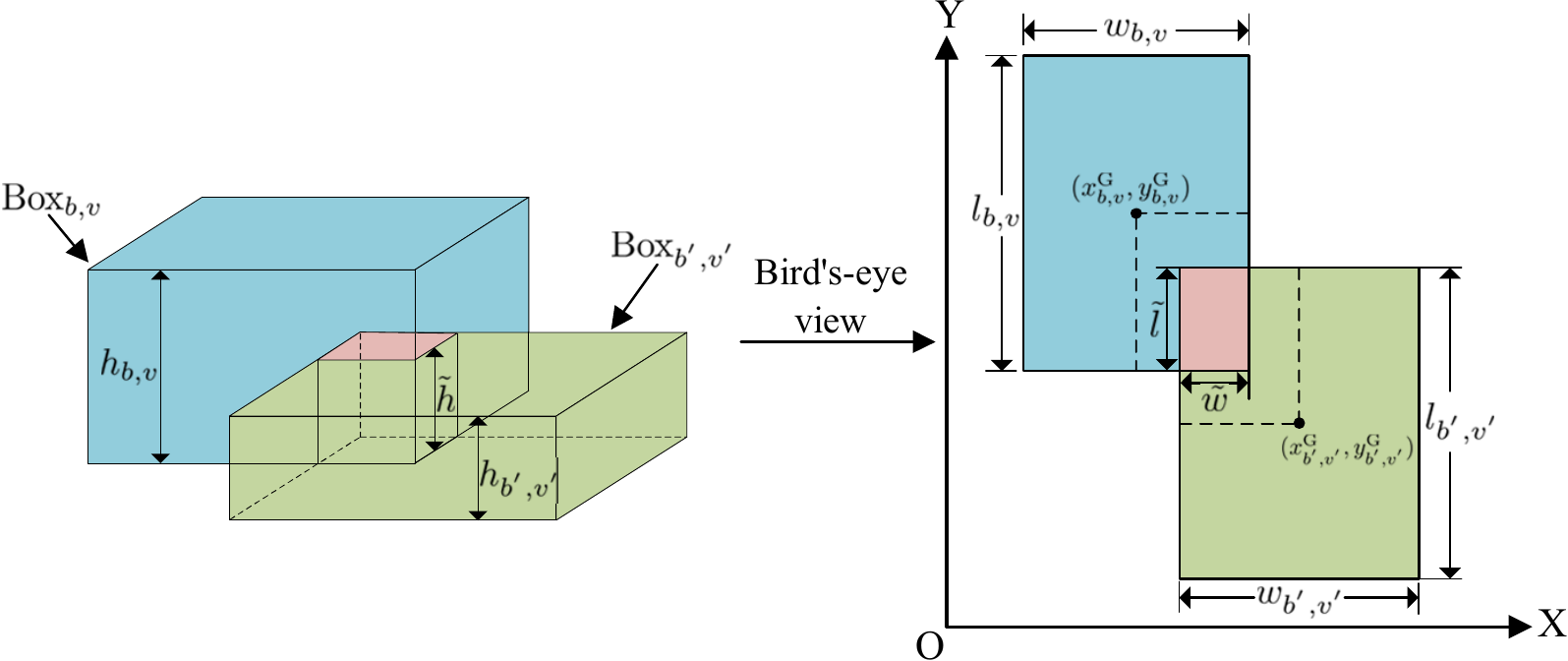}
		\caption{The bird's-eye view for the IoU of two different 3D BBoxes $\mathrm{Box}_{b,v}$ and $\mathrm{Box}_{b^{'},v^{'}}$.}
	\end{minipage}
\end{figure}

The tuple $\mathrm{Box}_{b,v}=(l_{b,v},w_{b,v},h_{b,v},x^{\mathrm{G}}_{b,v}, y^{\mathrm{G}}_{b,v}, z^{\mathrm{G}}_{b,v},\phi^{\mathrm{G}}_{b,v})$ can be used to express the 3D BBox of the $(b,v)$th vehicle, and all the obtained 3D BBoxes can build a set $\bm{\mathcal{X}}=\{\mathrm{Box}_{b,v}\ |\ v=1,2,\cdots,N_{\mathrm{V},b}, b\in\bm{\mathcal{B}}\}$. However, since the FOV of different cameras will inevitably overlap, the same vehicle is possible to be captured by different cameras, as shown in Fig.~5. Hence, a vehicle may be detected repeatedly from different images to generate multiple nearly identical 3D BBoxes in $\bm{\mathcal{X}}$. To remove the redundant 3D BBoxes, we select the intersecting BBoxes with the overlap degree exceeding a preset threshold for each vehicle and obtain the average of the selected BBoxes as the single BBox for the target vehicle.

Specifically, we utilize the intersection over union (IoU) to indicate the degree of the overlap of the two different 3D BBoxes, where the IoU is defined as the ratio of overlap volume to union volume of the two 3D BBoxes. Since all vehicles run on the ground and the azimuths of all vehicles are approximately parallel with the lane direction \cite{AKlautau},\cite{WXu1},\cite{YTian}-\cite{GCharan1}, i.e., the $\mathrm{Y}^{\mathrm{G}}$ axis, we can obtain $\phi^{\mathrm{G}}_{b,v}\approx 0$ or $\phi^{\mathrm{G}}_{b,v}\approx \pi$, and $z^{\mathrm{G}}_{b,v}\approx \frac{h_{b,v}}{2}$, $v=1,2,\cdots,N_{\mathrm{V},b}$, $b\in\bm{\mathcal{B}}$. Thus, as illustrated in Fig.~6, the height $\tilde{h}$ of the overlapping area of $\mathrm{Box}_{b,v}$ and $\mathrm{Box}_{b^{'},v^{'}}$ will be the minimum value of the heights of $\mathrm{Box}_{b,v}$ and $\mathrm{Box}_{b^{'},v^{'}}$, i.e., $\min(h_{b,v},h_{b^{'},v^{'}})$. The length $\tilde{l}$ and the width $\tilde{
w}$ of the overlapping area of $\mathrm{Box}_{b,v}$ and $\mathrm{Box}_{b^{'},v^{'}}$ can be obtained as $\min(w_{b,v},w_{b^{'},v^{'}},\frac{1} {2}w_{b,v}+\frac{1}{2}w_{b^{'},v^{'}}-|x^{\mathrm{G}}_{b,v}-x^{\mathrm{G}}_{b^{'},v^{'}}|)$ and $\min(l_{b,v},l_{b^{'},v^{'}},\frac{1}{2}l_{b,v}+\frac{1}{2}l_{b^{'},v^{'}}-|y^{\mathrm{G}}_{b,v}-y^{\mathrm{G}}_{b^{'},v^{'}}|)$, respectively. When $\mathrm{Box}_{b,v}$ and $\mathrm{Box}_{b^{'},v^{'}}$ do not overlap, the length and the width of the overlapping area are set as 0. Thus, the volume of the overlapping area is given by
\begin{equation}
\begin{aligned}
V_{\mathrm{O}}&=\max(0,\tilde{l})\max(0,\tilde{w})\tilde{h}\\
&
\begin{aligned}
=&\max(0,\min(w_{b,v},w_{b^{'},v^{'}},\frac{1} {2}w_{b,v}+\frac{1}{2}w_{b^{'},v^{'}}-|x^{\mathrm{G}}_{b,v}-x^{\mathrm{G}}_{b^{'},v^{'}}|))\\
&\cdot\max(0,\min(l_{b,v},l_{b^{'},v^{'}},\frac{1}{2}l_{b,v}+\frac{1}{2}l_{b^{'},v^{'}}-|y^{\mathrm{G}}_{b,v}-y^{\mathrm{G}}_{b^{'},v^{'}}|))\\
&\cdot \min(h_{b,v},h_{b^{'},v^{'}}).
\end{aligned}
\end{aligned}
\end{equation}
The IoU of $\mathrm{Box}_{b,v}$ and $\mathrm{Box}_{b^{'},v^{'}}$ can be expressed as
\begin{equation}
\mathrm{IoU}_{\mathrm{3D}}(b,v,b^{'},v^{'})=\frac{V_{\mathrm{O}}}{l_{b,v}w_{b,v}h_{b,v}+l_{b^{'},v^{'}}w_{b^{'},v^{'}}h_{b^{'},v^{'}}-V_{\mathrm{O}}}.
\end{equation}

Then, we randomly select a BBox from $\bm{\mathcal{X}}$ and denote it as $\mathrm{Box}_{b_{\mathrm{s}},v_{\mathrm{s}}}$. We next calculate the IoU of $\mathrm{Box}_{b_{\mathrm{s}},v_{\mathrm{s}}}$ and every other BBox in $\bm{\mathcal{X}}$ and gather the BBoxes whose IoU can exceed a preset threshold $\gamma$ to form a set $\bm{\mathcal{X}}_{\mathrm{s}}$, i.e.,
\begin{equation}
\bm{\mathcal{X}}_{\mathrm{s}}=\{\mathrm{Box}_{b,v}\ |\ \mathrm{IoU}_{\mathrm{3D}}(b,v,b_{\mathrm{s}},v_{\mathrm{s}})>\gamma,\ \forall\ \mathrm{Box}_{b,v}\in \tilde{\bm{\mathcal{X}}}\},
\end{equation}
where $\tilde{\bm{\mathcal{X}}}=\bm{\mathcal{X}}\backslash \{\mathrm{Box}_{b_{\mathrm{s}},v_{\mathrm{s}}}\}$. Denoting $\bm{\mathcal{X}}_{\mathrm{ave}}=\bm{\mathcal{X}}_{\mathrm{s}}\cup \{\mathrm{Box}_{b_{\mathrm{s}},v_{\mathrm{s}}}\}$, all the BBoxes in $\bm{\mathcal{X}}_{\mathrm{ave}}$ are merged to obtain an average BBox $\mathrm{Box}_{\mathrm{ave}}$, where the length, the width, the height, and the center location of $\mathrm{Box}_{\mathrm{ave}}$ are given by $l_{\mathrm{ave}}=\frac{1}{\mathrm{Card}(\bm{\mathcal{X}}_{\mathrm{ave}})}\sum_{\mathrm{Box}_{b,v}\in\bm{\mathcal{X}}_{\mathrm{ave}}}l_{b,v}$, $w_{\mathrm{ave}}=\frac{1}{\mathrm{Card}(\bm{\mathcal{X}}_{\mathrm{ave}})}\sum_{\mathrm{Box}_{b,v}\in\bm{\mathcal{X}}_{\mathrm{ave}}}w_{b,v}$, $h_{\mathrm{ave}}=\frac{1}{\mathrm{Card}(\bm{\mathcal{X}}_{\mathrm{ave}})}\sum_{\mathrm{Box}_{b,v}\in\bm{\mathcal{X}}_{\mathrm{ave}}}h_{b,v}$, and $(x^{\mathrm{G}}_{\mathrm{ave}}, y^{\mathrm{G}}_{\mathrm{ave}}, z^{\mathrm{G}}_{\mathrm{ave}})=\frac{1}{\mathrm{Card}(\bm{\mathcal{X}}_{\mathrm{ave}})}\sum_{\mathrm{Box}_{b,v}\in\bm{\mathcal{X}}_{\mathrm{ave}}}(x^{\mathrm{G}}_{b,v}, y^{\mathrm{G}}_{b,v}, z^{\mathrm{G}}_{b,v})$, respectively. Thus, all the BBoxes in $\bm{\mathcal{X}}_{\mathrm{ave}}$ can be replaced by $\mathrm{Box}_{\mathrm{ave}}$ to remove the redundancy of BBoxes in $\bm{\mathcal{X}}$. For the remaining BBoxes in $\bm{\mathcal{X}}$, we can repeat the above BBox merging operation to eliminate all the redundant BBoxes, and thereby obtain a de-redundancy BBox set $\bm{\mathcal{X}}_{\mathrm{E}}$ in which the seriously overlapped BBoxes are removed. The detailed steps of the BBox elimination method is shown in Algorithm 1.

\begin{algorithm}
  \caption{The BBox Elimination Method}
  \label{alg1}
  \begin{algorithmic}[1]
  \REQUIRE ~~\\ 
  The set $\bm{\mathcal{X}}$ and the preset threshold $\gamma<1$ of IoU;\\
  \ENSURE ~~\\ 
  The de-redundancy BBox set $\bm{\mathcal{X}}_{\mathrm{E}}$;
  \STATE Set $\bm{\mathcal{X}}_{\mathrm{E}}=\emptyset$;
  \REPEAT
  \STATE Randomly select a BBox from $\bm{\mathcal{X}}$ as $\mathrm{Box}_{b_{\mathrm{s}},v_{\mathrm{s}}}$;
  \STATE Calculate $\mathrm{IoU}_{\mathrm{3D}}(b,v,b_{\mathrm{s}},v_{\mathrm{s}})$, $\forall\ \mathrm{Box}_{b,v}\in \tilde{\bm{\mathcal{X}}}=\bm{\mathcal{X}}\backslash \{\mathrm{Box}_{b_{\mathrm{s}},v_{\mathrm{s}}}\}$;
  \STATE Obtain $\bm{\mathcal{X}}_{\mathrm{ave}}=\{\mathrm{Box}_{b,v}\ |\ \mathrm{IoU}_{\mathrm{3D}}(b,v,b_{\mathrm{s}},v_{\mathrm{s}})>\gamma,\ \forall\ \mathrm{Box}_{b,v}\in\bm{\mathcal{X}}\}$;
  \STATE Obtain the average BBox $\mathrm{Box}_{\mathrm{ave}}$ merged from the BBoxes in $\bm{\mathcal{X}}_{\mathrm{ave}}$ by calculating $l_{\mathrm{ave}}$, $w_{\mathrm{ave}}$, $h_{\mathrm{ave}}$ and $(x^{\mathrm{G}}_{\mathrm{ave}}, y^{\mathrm{G}}_{\mathrm{ave}}, z^{\mathrm{G}}_{\mathrm{ave}})$;
  \STATE Set $\bm{\mathcal{X}}_{\mathrm{E}}=\bm{\mathcal{X}}_{\mathrm{E}}\cup\{\mathrm{Box}_{\mathrm{ave}}\}$;
  \STATE Set $\bm{\mathcal{X}}=\bm{\mathcal{X}}\backslash \bm{\mathcal{X}}_{\mathrm{ave}}$;
  \UNTIL{$\bm{\mathcal{X}}=\emptyset$}.
  \end{algorithmic}
\end{algorithm}

\subsection{User Matching Stage}
After the 3D detection stage, the de-redundancy BBox set $\bm{\mathcal{X}}_{\mathrm{E}}$ contains the BBoxes of all vehicles. The BBox of each user in $\bm{\mathcal{X}}_{\mathrm{E}}$ needs to be matched with the radio signal of the corresponding BS. Here, we adopt the optimal beam pair between the BS and the user as the characteristic for user matching.

The user matching stage contains the following four steps:
\begin{itemize}
    \item[1)]
    Each BS collects the de-redundancy BBox sets extracted from the images at multiple moments and obtains the optimal beam pair index for each moment.
    \item[2)]
    Each BS performs grid division for each collected de-redundancy BBox set to obtain a BDF.
    \item[3)]
    We design a user matching DNN (UMAN) to infer a keypoint heatmap containing the user location information from both the BDF sequence and the beam pair sequence.
    \item[4)]
    Each BS estimates the location of the corresponding user by the keypoint heatmap and searches the BBox with the minimum distance from the estimated user location in the de-redundancy BBox set. The obtained BBox is matched with the user.
\end{itemize}

Specifically, the diagram of user matching stage is shown in Fig.~7. We consider the optimal beam pair will change slowly during the beam coherent time (BCT). Thus, all the serving BSs will perform beam alignment within each BCT to obtain the optimal beam pair. We denote the shooting interval of each camera as $T_{\mathrm{f}}$. For the $B$ images taken at the moment $k\alpha T_{\mathrm{f}}$, we implement the 3D detection method and BBox elimination method to obtain the de-redundancy BBox set $\bm{\mathcal{X}}_{\mathrm{E},k}$, $k=0,1,2,\cdots$, where $\alpha$ is an adjustable positive integer to control the frequency of 3D detection. Furthermore, for the $b_u$th BS and the $u$th user, we denote the index of the optimal beam pair obtained at the BCT containing the moment $ k\alpha T_{\mathrm{f}} $ as $I_{u,k}$, where the beam pair is selected from the beam pair set $\bm{\mathcal{P}}=\{(\bm{\mathrm{f}}^{\mathrm{B},1},\bm{\mathrm{f}}^{\mathrm{U},1}),(\bm{\mathrm{f}}^{\mathrm{B},1},\bm{\mathrm{f}}^{\mathrm{U},2}),\cdots,(\bm{\mathrm{f}}^{\mathrm{B},N_\mathrm{CB}},\bm{\mathrm{f}}^{\mathrm{U},N_{\mathrm{CU}}})\}$.

\begin{figure}[t]
	\begin{minipage}[t]{0.5\linewidth}
		\centering
			\includegraphics[width=90mm]{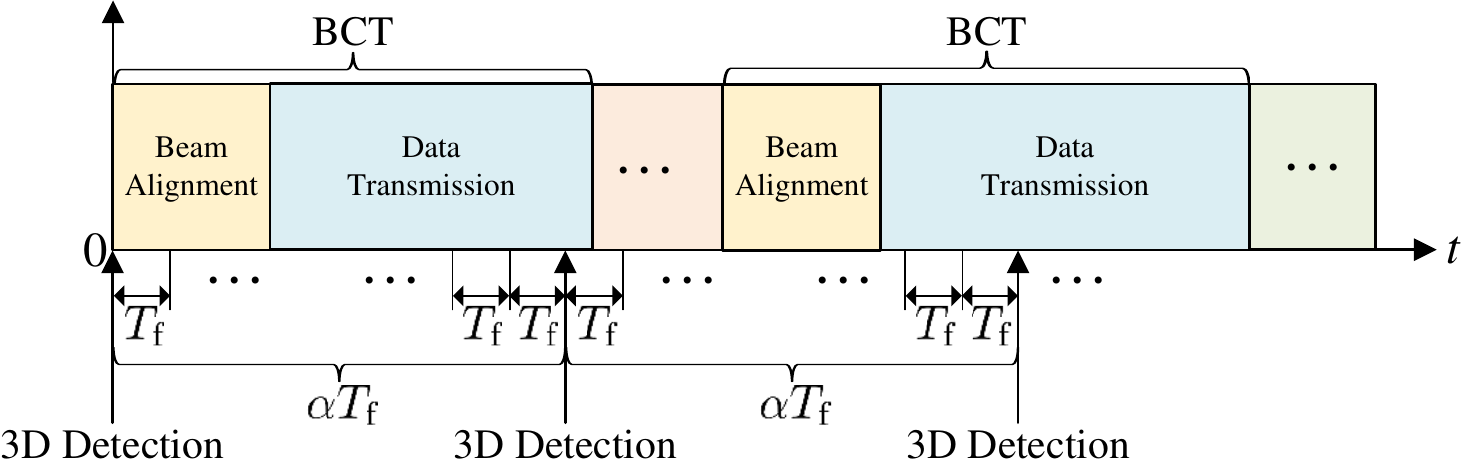}
		\caption{The diagram for user matching. Each BCT includes beam alignment and data transmission.}
	\end{minipage}
	\hspace{1ex}
	\begin{minipage}[t]{0.5\linewidth}
		\centering
	\includegraphics[width=50mm]{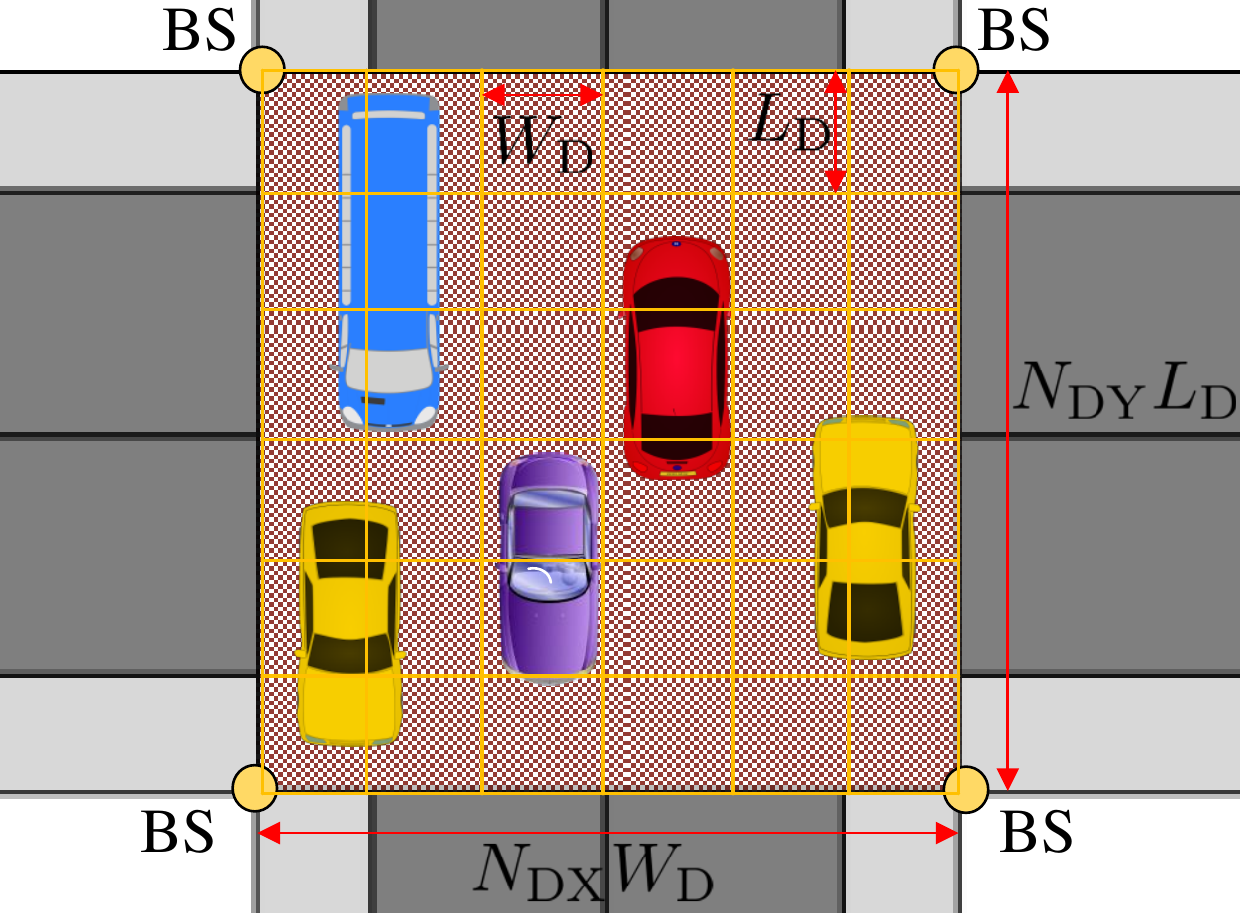}
		\caption{The grid division for the crossroad. The length and width of each grid are $L_{\mathrm{D}}$ and $W_{\mathrm{D}}$.}
	\end{minipage}
\end{figure}

Thus, the $b_u$th BS can collect the visual data sequence $\bm{\mathcal{X}}_{\mathrm{E},k}$ from central processing unit, $k=0,1,2,\cdots$, and the beam pair sequence $I_{u,k}$, $k=0,1,2,\cdots$. The matching task for each user can be realized only by the corresponding serving BS, since the optimal beam pair between the user and the corresponding serving BS is independent of other BSs. Therefore, for each serving BS, we design a UMAN to identify the BBox of the corresponding user with the aid of the collected sequence of visual data and beams. For clarity, we will illustrate UMAN under the scenario with a single BS and a single user, and thereby omit the subscript $u$ in the following discussion.

We utilize $\bm{\mathcal{X}}_{\mathrm{E},k}$ to design a BDF $\bm{D}_k$ as a part of inputs of UMAN. The BDF sequence can provide the spatial-temporal distribution information of the BBoxes of all vehicles in the communications environment for UMAN. The BDF is designed by performing grid division on the crossroad. For each grid, we use a vector to represent the average size of the BBoxes contained in the grid, and the vectors of all grids are stacked to generate the BDF.

Specifically, the plane of the crossroad, i.e., the $\mathrm{X}^{\mathrm{G}}-\mathrm{Y}^{\mathrm{G}}$ plane, is divided into the grids with equal size, as shown in Fig.~8. The grid length and the grid width are denoted as $L_{\mathrm{D}}$ and $W_{\mathrm{D}}$, respectively. Moreover, we denote the number of columns and rows of the grids that intersect with the crossroad as $N_{\mathrm{DX}}$ and $N_{\mathrm{DY}}$ respectively and denote the grid of the $n_{\mathrm{DX}}$th column and the $n_{\mathrm{DY}}$th row as the $(n_{\mathrm{DX}},n_{\mathrm{DY}})$th grid, $n_{\mathrm{DX}}=1,2,\cdots,N_{\mathrm{DX}}$, $n_{\mathrm{DY}}=1,2,\cdots,N_{\mathrm{DY}}$. Then, we select the BBoxes whose $\mathrm{X}^{\mathrm{G}}-\mathrm{Y}^{\mathrm{G}}$ plane locations are inside the $(n_{\mathrm{DX}},n_{\mathrm{DY}})$th grid from $\bm{\mathcal{X}}_{\mathrm{E},k}$ to form a set $\bm{\mathcal{X}}_{\mathrm{E},k}^{n_{\mathrm{DX}},n_{\mathrm{DY}}}$, where
\begin{equation}
\begin{aligned}
\bm{\mathcal{X}}_{\mathrm{E},k}^{n_{\mathrm{DX}},n_{\mathrm{DY}}}=\{\mathrm{Box}_{\mathrm{E}}\ |\ x^{\mathrm{G}}_{\mathrm{V},n_{\mathrm{DX}},n_{\mathrm{DY}}}\leq &x_{\mathrm{E}}^{\mathrm{G}}<x^{\mathrm{G}}_{\mathrm{V},n_{\mathrm{DX}},n_{\mathrm{DY}}}+W_{\mathrm{D}},\ \\
&y^{\mathrm{G}}_{\mathrm{V},n_{\mathrm{DX}},n_{\mathrm{DY}}}\leq y_{\mathrm{E}}^{\mathrm{G}}<y^{\mathrm{G}}_{\mathrm{V},n_{\mathrm{DX}},n_{\mathrm{DY}}}+L_{\mathrm{D}},\ \forall\ \mathrm{Box}_{\mathrm{E}}\in \bm{\mathcal{X}}_{\mathrm{E},k}\},
\end{aligned}
\end{equation}
$(x_{\mathrm{E}}^{\mathrm{G}},y_{\mathrm{E}}^{\mathrm{G}})$ is the plane coordinates of the center location of $\mathrm{Box}_{\mathrm{E}}$ and $(x^{\mathrm{G}}_{\mathrm{V},n_{\mathrm{DX}},n_{\mathrm{DY}}},y^{\mathrm{G}}_{\mathrm{V},n_{\mathrm{DX}},n_{\mathrm{DY}}})$ is the plane coordinates of the bottom-left vertex of the $(n_{\mathrm{DX}},n_{\mathrm{DY}})$th grid. The BBox $\mathrm{Box}_{\mathrm{ave}}^{n_{\mathrm{DX}},n_{\mathrm{DY}}}$ is obtained by averaging the length, width, height, and center location of the BBoxes in $\bm{\mathcal{X}}_{\mathrm{E},k}^{n_{\mathrm{DX}},n_{\mathrm{DY}}}$. We denote the length, width, and height of $\mathrm{Box}_{\mathrm{ave}}^{n_{\mathrm{DX}},n_{\mathrm{DY}}}$ as $l_{\mathrm{ave}}^{n_{\mathrm{DX}},n_{\mathrm{DY}}}$, $w_{\mathrm{ave}}^{n_{\mathrm{DX}},n_{\mathrm{DY}}}$, and $h_{\mathrm{ave}}^{n_{\mathrm{DX}},n_{\mathrm{DY}}}$ respectively and denote the maximum length, width, and height of all possible vehicle types as $l_{\mathrm{max}}$, $w_{\mathrm{max}}$, and $h_{\mathrm{max}}$, respectively. The BDF $\bm{D}_k$ is designed as an $N_{\mathrm{DX}}\times N_{\mathrm{DY}}\times 3$ dimensional tensor, and the $(n_{\mathrm{DX}},n_{\mathrm{DY}})$th row of $\bm{D}_k$ is set as the normalized BBox size $[\frac{l_{\mathrm{ave}}^{n_{\mathrm{DX}},n_{\mathrm{DY}}}}{l_{\mathrm{max}}},\frac{w_{\mathrm{ave}}^{n_{\mathrm{DX}},n_{\mathrm{DY}}}}{w_{\mathrm{max}}},\frac{h_{\mathrm{ave}}^{n_{\mathrm{DX}},n_{\mathrm{DY}}}}{h_{\mathrm{max}}}]$. For the grid that does not contain any BBoxes, the corresponding row of $\bm{D}_k$ is set as the zero vector. In general, the grid size should be small to reduce the possibility that a grid contains multiple BBoxes, i.e., $\mathrm{Card}(\bm{\mathcal{X}}_{\mathrm{E},k}^{n_{\mathrm{DX}},n_{\mathrm{DY}}})>1$, and thus, the BBox distribution information loss caused by averaging operation can be decreased.

Therefore, the inputs of UMAN should include the BDF sequence $\bm{D}_{k-M+1},\bm{D}_{k-M+2},\cdots,\bm{D}_k$ and beam pair sequence $I_{k-M+1},I_{k-M+2},\cdots,I_{k}$, $k=M-1,M,\cdots$, where $M$ is the input sequence length. Next, we analyze how to reasonably design the representation form of the output of UMAN. If each BBox in $\bm{\mathcal{X}}_{\mathrm{E},k}$ is regarded as a category, then the task of identifying the BBox of the user will be essentially a multi-class classification problem, as indicated in \cite{VMPinho}. However, the number of BBoxes in $\bm{\mathcal{X}}_{\mathrm{E},k}$ is random in practice, which makes the number of categories vary. Thus, the generally used one-hot encoding for representing the class label is not suitable as the output of UMAN.

\begin{figure}[t]
	\begin{minipage}[t]{0.5\linewidth}
		\centering
	\includegraphics[width=93mm]{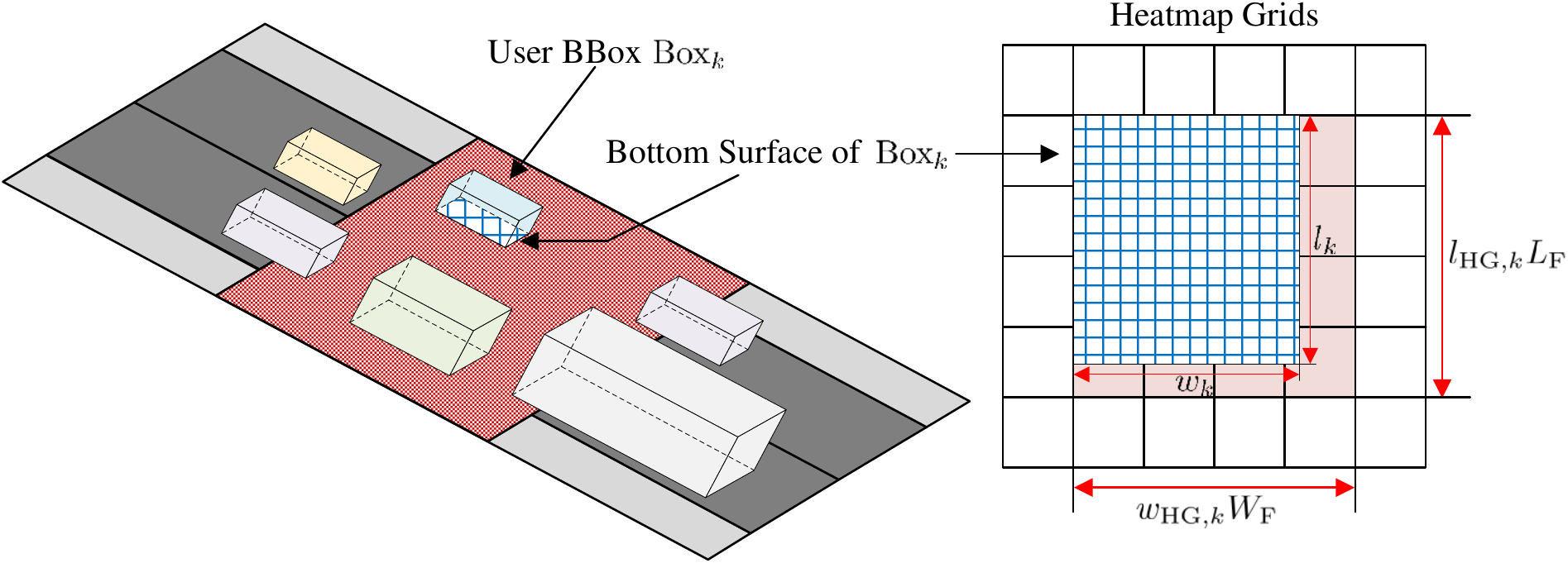}
		\caption{The bottom surface of user BBox $\mathrm{Box}_k$. The bottom surface occupies at least $l_{\mathrm{HG}, k}\times w_{\mathrm{HG}, k}$ heatmap grids.}
	\end{minipage}
	\hspace{1ex}
	\begin{minipage}[t]{0.5\linewidth}
		\centering
			\includegraphics[width=50mm]{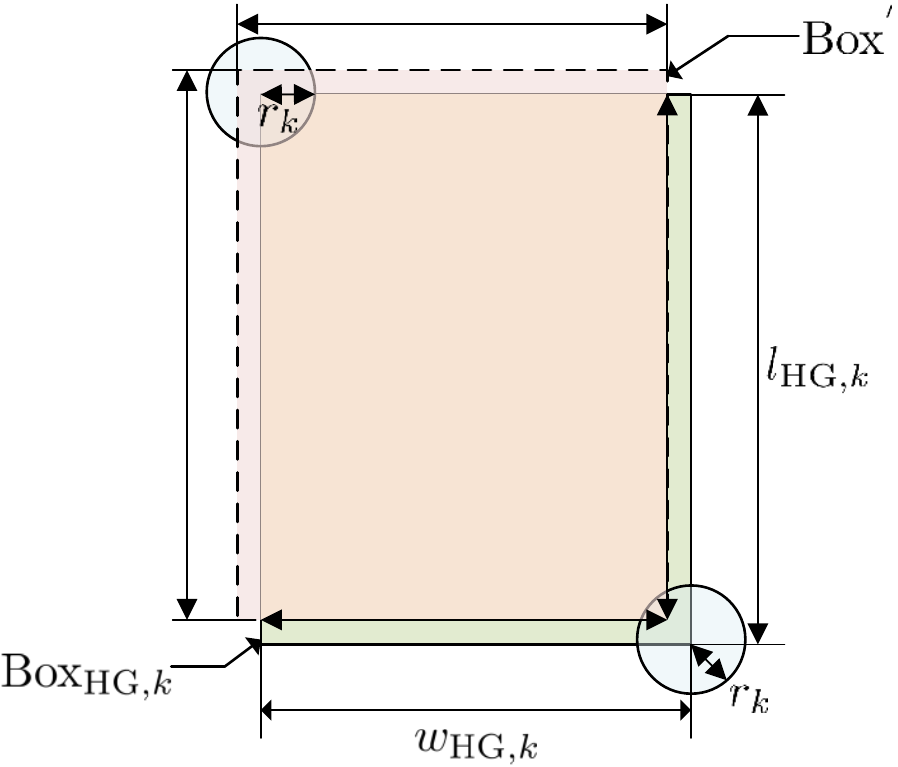}
		\caption{The BBox $\mathrm{Box}^{'}$ generated by the circle area of two corner points of $\mathrm{Box}_{\mathrm{HG},k}$. The radius of the circle area is $r_{k}$.}
	\end{minipage}
\end{figure}

The keypoint heatmap \cite{XZhou} can effectively represent the user location by the grid form, and the BBox of the user can be obtained by searching the BBox with the minimum distance from the user location in $\bm{\mathcal{X}}_{\mathrm{E},k}$. Hence, we utilize the keypoint heatmap as the output of UMAN. Specifically, we still divide the plane of crossroad into the grids of equal size. For clarity, the grids used for generating the keypoint heatmap are named as \emph{heatmap grids} to distinguish from the grids used for generating BDF. The length and the width of each heatmap grid are denoted as $L_{\mathrm{F}}$ and $W_{\mathrm{F}}$, respectively. We denote the number of columns and rows of the heatmap grids that intersect with the crossroad as $N_{\mathrm{FX}}$ and $N_{\mathrm{FY}}$ respectively and denote the BBox in $\bm{\mathcal{X}}_{\mathrm{E},k}$ of the user as $\mathrm{Box}_{k}$. Thus, the keypoint heatmap is designed as an $N_{\mathrm{FX}}\times N_{\mathrm{FY}}$ dimensional tensor $\bm{F}_{k}$. We use a two-dimensional Gaussian kernel to represent $\mathrm{Box}_{k}$ onto the heatmap $\bm{F}_{k}$. The index of the heatmap grid that contains the plane coordinates of the center location of $\mathrm{Box}_{k}$ is obtained as $(n_{\mathrm{X},k},n_{\mathrm{Y},k})$. Then, the Gaussian kernel is expressed as
\begin{equation}
G_{k}(n_{\mathrm{FX}},n_{\mathrm{FY}})=e^{-\frac{(n_{\mathrm{FX}}-n_{\mathrm{X},k})^2+(n_{\mathrm{FY}}-n_{\mathrm{Y},k})^2}{2\sigma_{k}^2}},
\end{equation}
where $\sigma_{k}$ is the standard deviation depending on the length $l_{k}$ and the width $w_{k}$ of $\mathrm{Box}_{k}$. We here adopt the widely used scheme in CornerNet \cite{HLaw} to determine $\sigma_{k}$. As shown in Fig.~9, the bottom surface of $\mathrm{Box}_{k}$ occupies at least $l_{\mathrm{HG}, k}\times w_{\mathrm{HG}, k}$ heatmap grids, where $l_{\mathrm{HG}, k}=\left\lceil \frac{l_{k}}{L_{\mathrm{F}}}\right\rceil$ and $w_{\mathrm{HG}, k}=\left\lceil \frac{w_{k}}{W_{\mathrm{F}}}\right\rceil$. We denote the heatmap grid area occupied by the bottom surface of $\mathrm{Box}_{k}$ as the 2D BBox $\mathrm{Box}_{\mathrm{HG},k}$ on the heatmap. The length and the width of $\mathrm{Box}_{\mathrm{HG},k}$ are $l_{\mathrm{HG},k}$ and $w_{\mathrm{HG},k}$, respectively. As shown in Fig.~10, the standard deviation $\sigma_{k}$ is given by a radius $r_{k}$. Note that $r_{k}$ should guarantee that $\mathrm{Box}_{\mathrm{HG},k}$ and any 2D BBox $\mathrm{Box}^{'}$ can achieve at least $\tilde{\gamma}$ IoU, where the two corner points of $\mathrm{Box}^{'}$ should be in the circle areas around the two corner points of $\mathrm{Box}_{\mathrm{HG},k}$. Thus, the radius $r_{k}$ can be given by $r_{k}=\min(r_1,r_2,r_3)$ \cite{GaussianRadius}, where
\begin{equation}
\begin{aligned}
r_1&=\frac{(l_{\mathrm{HG},k}+w_{\mathrm{HG},k})-\sqrt{(l_{\mathrm{HG},k}+w_{\mathrm{HG},k})^2-4l_{\mathrm{HG},k} w_{\mathrm{HG},k}\frac{(1-\tilde{\gamma})}{(1+\tilde{\gamma})}}}{2},\\
r_2&=\frac{(l_{\mathrm{HG},k}+w_{\mathrm{HG},k})-\sqrt{(l_{\mathrm{HG},k}+w_{\mathrm{HG},k})^2-4l_{\mathrm{HG},k}w_{\mathrm{HG},k}(1-\tilde{\gamma})}}{4},\\
r_3&=\frac{-\tilde{\gamma}(l_{\mathrm{HG},k}+w_{\mathrm{HG},k})+\sqrt{\tilde{\gamma}^2(l_{\mathrm{HG},k}+w_{\mathrm{HG},k})^2-4l_{\mathrm{HG},k}w_{\mathrm{HG},k}\tilde{\gamma}(1-\tilde{\gamma})}}{4\tilde{\gamma}}.
\end{aligned}
\end{equation}
The standard deviation $\sigma_{k}$ is determined as $\frac{2\lfloor r_{k}\rfloor+1}{6}$, and the heatmap $\bm{F}_{k}$ is designed as
\begin{equation}
\bm{F}_{k}[n_{\mathrm{FX}},n_{\mathrm{FY}}]=\left\{ \begin{aligned}
&G_{k}(n_{\mathrm{FX}},n_{\mathrm{FY}}),\
\begin{aligned}
&n_{\mathrm{FX}}\in \{n_{\mathrm{X},k}-\lfloor r_{k}\rfloor,n_{\mathrm{X},k}-\lfloor r_{k}\rfloor+1,\cdots,n_{\mathrm{X},k}+\lfloor r_{k}\rfloor\},\\ &n_{\mathrm{FY}}\in \{n_{\mathrm{Y},k}-\lfloor r_{k}\rfloor,n_{\mathrm{Y},k}-\lfloor r_{k}\rfloor+1,\cdots,n_{\mathrm{Y},k}+\lfloor r_{k}\rfloor\},\\
&1\leq n_{\mathrm{FX}}\leq N_{\mathrm{FX}}, 1\leq n_{\mathrm{FY}}\leq N_{\mathrm{FY}}
\end{aligned}
\\
&0,\ \mathrm{otherwise}.
\end{aligned}\right.
\end{equation}
From equations (9) and (10), $\bm{F}_{k}$ will contain the effective distribution information of the location of $\mathrm{Box}_{k}$. Once we obtain the estimate $\hat{\bm{F}}_{k}$ through UMAN, the index $(n^{\mathrm{max}}_{\mathrm{X},k},n^{\mathrm{max}}_{\mathrm{Y},k})$ of the maximum element of $\hat{\bm{F}}_{k}$, i.e., the peak point of the Gaussian kernel, can be used to indicate the plane location of the $\mathrm{Box}_{k}$.
Specifically, the plane location of $\mathrm{Box}_{k}$ can be estimated as the center location $(\hat{x}_{\mathrm{ES},k},\hat{y}_{\mathrm{ES},k})$ of the $(n^{\mathrm{max}}_{\mathrm{X},k},n^{\mathrm{max}}_{\mathrm{Y},k})$th heatmap grid. Therefore, the $\mathrm{Box}_{k}$ can be obtained by selecting the BBox whose plane location is closest to $(\hat{x}_{\mathrm{ES},k},\hat{y}_{\mathrm{ES},k})$ in $\bm{\mathcal{X}}_{\mathrm{E},k}$.

\begin{figure}[t]
\centering
\includegraphics[width=0.91\textwidth]{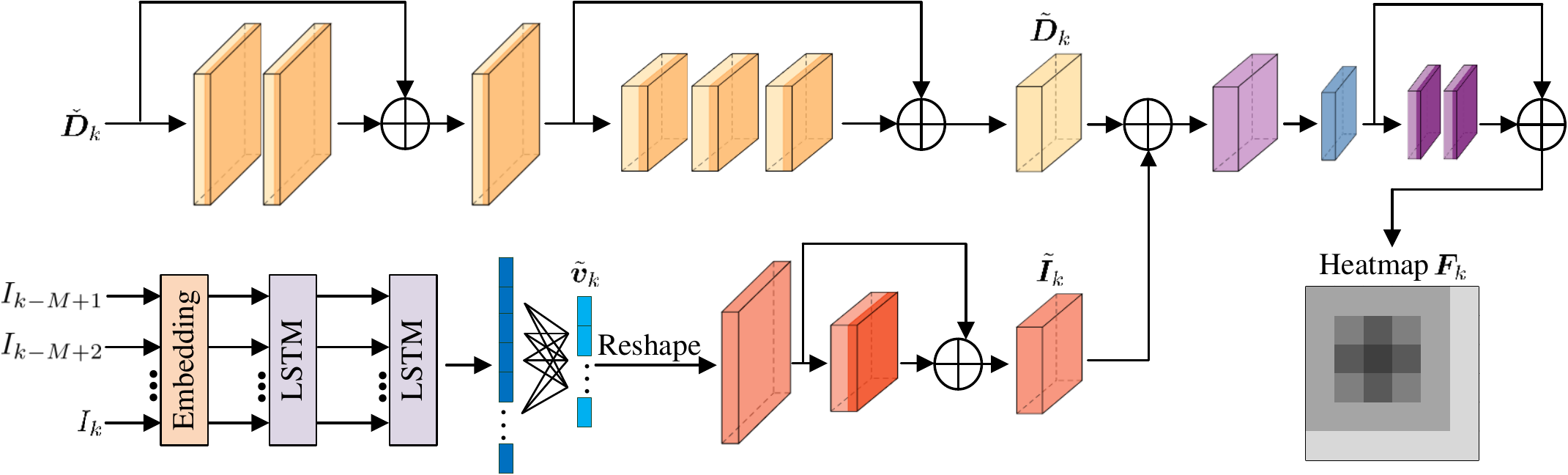}
\caption{The network structure of the proposed UMAN for the mapping from the BDF sequence $\bm{D}_{k-M+1},\bm{D}_{k-M+2},\cdots,\bm{D}_k$ and beam sequence $I_{k-M+1},I_{k-M+2},\cdots,I_{k}$ to the heatmap $\bm{F}_{k}$. The BDF sequence $\bm{D}_{k-M+1},\bm{D}_{k-M+2},\cdots,\bm{D}_k$ is stacked to generate $\check{\bm{D}}_{k}$.}
\end{figure}

We further design the network structure of UMAN to realize the mapping from the BDF sequence $\bm{D}_{k-M+1},\bm{D}_{k-M+2},\cdots,\bm{D}_k$ and the beam sequence $I_{k-M+1},I_{k-M+2},\cdots,I_{k}$ to the heatmap $\bm{F}_{k}$, $k=M-1,M,\cdots$. As shown in Fig.~11, the feature fusion architecture is utilized to build UMAN. Specifically, two different subnetworks are used for feature extraction of BDF sequence and beam sequence, respectively. For the subnetwork corresponding to the BDF sequence, we stack $\bm{D}_{k-M+1},\bm{D}_{k-M+2},\cdots,\bm{D}_k$ to generate an $N_{\mathrm{DX}}\times N_{\mathrm{DY}}\times 3M$ dimensional tensor $\check{\bm{D}}_{k}$, where $\check{\bm{D}}_{k}[:,:,(3m-2):3m]=\bm{D}_{k-M+m}$, $m=1,2,\cdots,M$. We then feed $\check{\bm{D}}_{k}$ into several 2D convolutional layers with the residual connection \cite{KHe} to obtain a tensor $\tilde{\bm{D}}_{k}$. For the subnetwork corresponding to the beam sequence, the beam sequence $I_{k-M+1},I_{k-M+2},\cdots,I_{k}$ is input into the embedding layer and the long short-term memory (LSTM) layers \cite{SHochreiter}. The output tensor of the final time step of the last LSTM layer will
be fed into the fully connected layers to obtain a vector $\tilde{\bm{v}}_{k}$ with the length $N_{\mathrm{DX}}N_{\mathrm{DY}}$. Next, $\tilde{\bm{v}}_{k}$ is reshaped into an $N_{\mathrm{DX}}\times N_{\mathrm{DY}}\times 1$ dimensional tensor and will be fed into several 2D convolutional layers to obtain a tensor $\tilde{\bm{I}}_{k}$ with the same dimension of $\tilde{\bm{D}}_{k}$. Then, the sum of tensor $\tilde{\bm{I}}_{k}$ and $\tilde{\bm{D}}_{k}$ is obtained for feature fusion and the obtained fusion tensor is input into the pooling layers and 2D convolutional layers to output $\bm{F}_{k}$, where the last convolutional layer adopts the Sigmoid activation function to ensure the range of all output values is between 0 and 1.

The above designed UMAN can be equipped for each BS. Once the UMANs of all BSs are well-trained, the $b_u$th BS can utilize its own UMAN at the moment $kT_{\mathrm{b}}$ to identify the BBox of the $u$th user.

\section{Vision Based Resource Allocation}
In mmWave network with multiple BSs and users, the power allocation and user scheduling are essential to reduce the interference from different BSs and to improve the total transmission rate of communications system \cite{LLiang}, as shown in Fig.~12. Specifically, for the $u$th user, the index $b_u$ of the connected BS and the transmission power $P_{u}$ should be adjusted to maximize the total transmission rate
\begin{equation}
R_{\mathrm{total}}=\sum_{u\in\bm{\mathcal{U}} }\log_2\left(1+\frac{P_{u}|(\bm{\mathrm{f}}^{\mathrm{U},\mathrm{opt}}_{b_u,u})^{\mathrm{H}}\bm{H}_{b_u,u}\bm{\mathrm{f}}^{\mathrm{B},\mathrm{opt}}_{b_u,u}|^2}{\sum_{u^{'}\in \bm{\mathcal{U}}\backslash\{u\}}P_{u^{'}}|(\bm{\mathrm{f}}^{\mathrm{U},\mathrm{opt}}_{b_u,u})^{\mathrm{H}}\bm{H}_{b_{u^{'}},u}\bm{\mathrm{f}}^{\mathrm{B},\mathrm{opt}}_{b_{u^{'}},u^{'}}|^2+\sigma^2}
\right).
\end{equation}
Thus, the resource allocation problem can be formulated as
\begin{subequations}
\begin{align}
\mathrm{P1}: &\max_{\bm{b},\bm{P}}\ R_{\mathrm{total}} \label{Za}\\
s.t.\ &\mathrm{C1}:\ \sum_{u\in\bm{\mathcal{U}}}\mathbbm{1}(b_u=b)=0\ \mathrm{or}\ 1, \forall b\in\bm{\mathcal{B}} \label{Zb},\\
&\mathrm{C2}:\ P_u\leq P_{\mathrm{max},b_u}, \forall u\in \bm{\mathcal{U}} \label{Zc},
\end{align}
\end{subequations}
where $\bm{b}=[b_1,b_2,\cdots,b_U]$, $\bm{P}=[P_1,P_2,\cdots,P_U]$, while the indicator function $\mathbbm{1}(\mathcal{C})$ is set as 1 when the condition $\mathcal{C}$ is true, or is 0 otherwise. The constraint (12b) guarantees that at most one user is connected to one BS, while the constraint (12c) ensures that the transmission power of the $b$th BS will not exceed the maximum transmission power $P_{\mathrm{max},b}$.

\begin{figure}[t]
\centering
\includegraphics[width=0.5\textwidth]{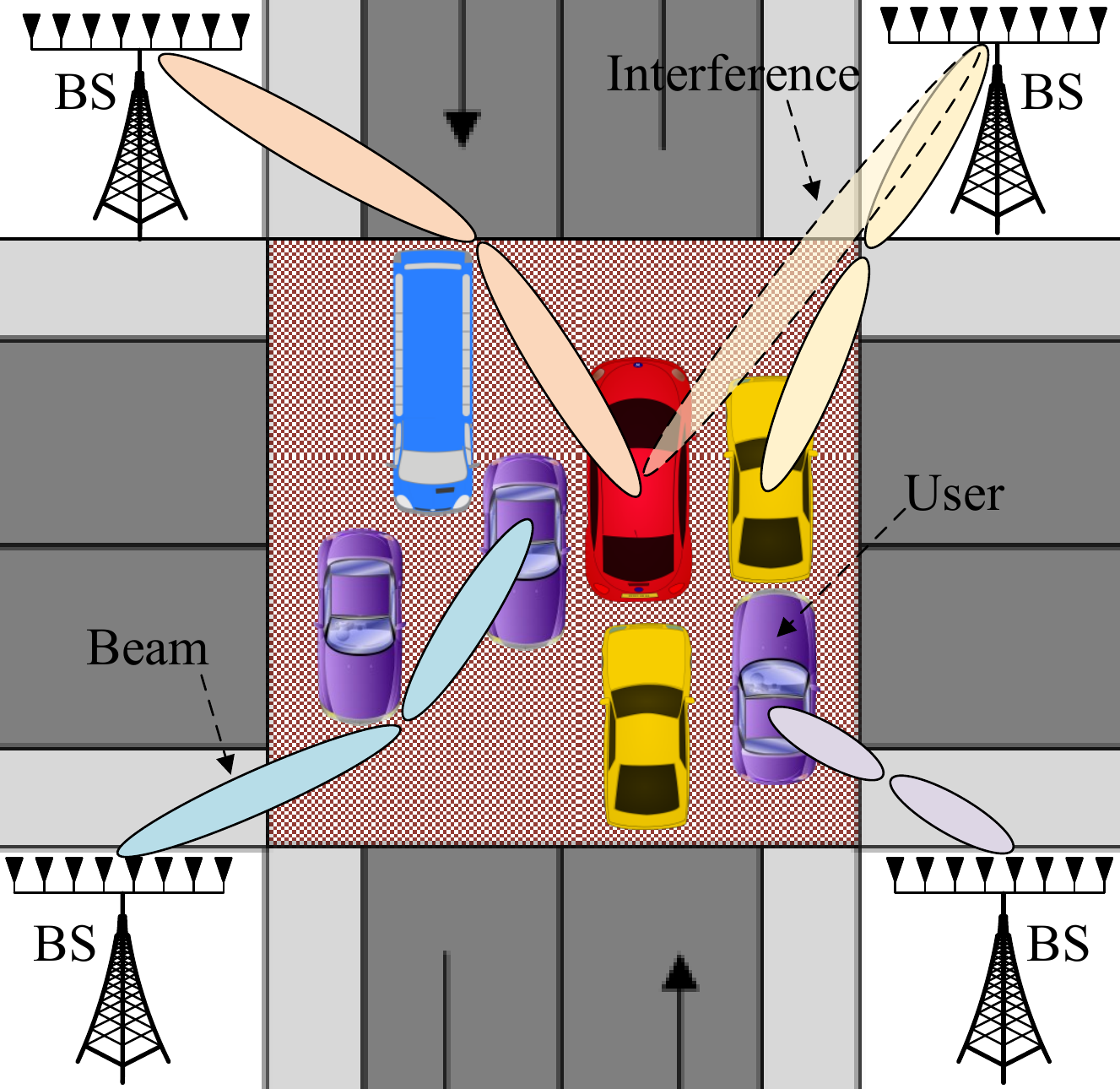}
\caption{User scheduling and power control under the scenario with multiple BSs and users. The transmission link between a BS and a user can be affected by the interference from other BSs.}
\end{figure}

The traditional resource allocation methods for P1 need the channel information between all BSs and users, i.e., $\bm{H}_{b,u}$, $u\in\bm{\mathcal{U}}$, $b\in\bm{\mathcal{B}}$, and thereby have inevitable time and spectrum overhead. Moreover, to solve P1, the conventional methods generally adopt the iterative optimization or the exhaustive search, and have too much computing overhead.

Hence, we propose a new method named VBRAM to predict the optimal $\bm{b}^{*}=[b_1^{*},b_2^{*},\cdots,b_U^{*}]$ and $\bm{P}^{*}=[P_1^{*},P_2^{*},\cdots,P_U^{*}]$ of P1 from environmental images. Since the BBox of each user can be identified by 3DUMM and since the non-user vehicles act as the scatters, we can distinguish the users' BBoxes and scatters' BBoxes in the de-redundancy BBox set. The obtained spatial distribution of BBoxes of all users as well as the other scattering objects can effectively reflect the characteristics and the correlation of the channels of all users. Therefore, VBRAM utilizes the spatial distribution information of the users and scattering objects to design the input feature of the resource allocation DNN, and thereby could predict $\bm{b}^*$ and $\bm{P}^*$ purely based on vision. Moreover, VBRAM can be low-latency, since the object distribution can be rapidly obtained by the object detection/tracking techniques \cite{Zliu}, and the inference speed of DNN is significantly fast with the aid of GPU.

Specifically, at the moment $kT_{\mathrm{b}}$, we utilize $\bm{\mathcal{X}}_{\mathrm{E},k}$ to design the USDF that is used as the input of vision based resource allocation DNN (VRAN) to predict the P1's optimal solution for the moment $kT_{\mathrm{b}}$. The USDF is also designed based on grid division and utilizes the number of users' BBoxes contained in each grid to differentiate the spatial distribution of users and other scattering objects. We divide the crossroad plane into $N_{\mathrm{RX}}\times N_{\mathrm{RY}}$ grids with equal size, where $N_{\mathrm{RX}}$ and $N_{\mathrm{RY}}$ are the number of columns and rows of the grids that intersect with the crossroad, respectively. For clarity, the grids used for producing USDF are called as \emph{USDF grids} to distinguish from all the other grids. We denote the length and the width of the USDF grid as $L_{\mathrm{R}}$ and $W_{\mathrm{R}}$ respectively and denote the BBox of the $u$th user in $\bm{\mathcal{X}}_{\mathrm{E},k}$ as $\mathrm{Box}_{k}^{u}$, $u\in\bm{\mathcal{U}}$. For the $(n_{\mathrm{RX}},n_{\mathrm{RY}})$th grid, $n_{\mathrm{RX}}=1,2,\cdots,N_{\mathrm{RX}}$, $n_{\mathrm{RY}}=1,2,\cdots,N_{\mathrm{RY}}$, we build a set $\bm{\mathcal{X}}_{\mathrm{RE},k}^{n_{\mathrm{RX}},n_{\mathrm{RY}}}$ by counting which BBoxes in $\bm{\mathcal{X}}_{\mathrm{E},k}$ are inside this grid. We calculate a new parameter $\mathrm{Box}_{\mathrm{R,ave}}^{n_{\mathrm{RX}},n_{\mathrm{RY}}}$ by averaging the BBoxes in $\bm{\mathcal{X}}_{\mathrm{RE},k}^{n_{\mathrm{RX}},n_{\mathrm{RY}}}$. Furthermore, we count the number $i^{n_{\mathrm{RX}},n_{\mathrm{RY}}}$ of the BBoxes of users in $\bm{\mathcal{X}}_{\mathrm{RE},k}^{n_{\mathrm{RX}},n_{\mathrm{RY}}}$, where $i^{n_{\mathrm{RX}},n_{\mathrm{RY}}}=\mathrm{Card}(\bm{\mathcal{U}}_{k}^{n_{\mathrm{RX}},n_{\mathrm{RY}}})$ and
\begin{equation}
\bm{\mathcal{U}}_{k}^{n_{\mathrm{RX}},n_{\mathrm{RY}}}=\{u\ |\ \mathrm{Box}_{k}^{u}\in\bm{\mathcal{X}}_{\mathrm{RE},k}^{n_{\mathrm{RX}},n_{\mathrm{RY}}},u\in\bm{\mathcal{U}}\}.
\end{equation}
Thus, the USDF is designed as an $N_{\mathrm{RX}}\times N_{\mathrm{RY}}\times 4$ tensor $\bm{Z}_{k}$. For the $(n_{\mathrm{RX}},n_{\mathrm{RY}})$th USDF grid with $i^{n_{\mathrm{RX}},n_{\mathrm{RY}}}>0$, the $(n_{\mathrm{RX}},n_{\mathrm{RY}})$th row of $\bm{Z}_{k}$ is set as $[\frac{l_{\mathrm{R,ave}}^{n_{\mathrm{RX}},n_{\mathrm{RY}}}}{l_{\mathrm{max}}},\frac{w_{\mathrm{R,ave}}^{n_{\mathrm{RX}},n_{\mathrm{RY}}}}{w_{\mathrm{max}}},\frac{h_{\mathrm{R,ave}}^{n_{\mathrm{RX}},n_{\mathrm{RY}}}}{h_{\mathrm{max}}},i^{n_{\mathrm{RX}},n_{\mathrm{RY}}}]$, where $l_{\mathrm{R,ave}}^{n_{\mathrm{RX}},n_{\mathrm{RY}}}$, $w_{\mathrm{R,ave}}^{n_{\mathrm{RX}},n_{\mathrm{RY}}}$ and $h_{\mathrm{R,ave}}^{n_{\mathrm{RX}},n_{\mathrm{RY}}}$ are the length, width, and height of $\mathrm{Box}_{\mathrm{R,ave}}^{n_{\mathrm{RX}},n_{\mathrm{RY}}}$, respectively. If $i^{n_{\mathrm{RX}},n_{\mathrm{RY}}}=0$ and $\mathrm{Card}(\bm{\mathcal{X}}_{\mathrm{RE},k}^{n_{\mathrm{RX}},n_{\mathrm{RY}}})\neq 0$, then the $(n_{\mathrm{RX}},n_{\mathrm{RY}})$th USDF grid contains the vehicles that act as scattering objects but not the users. Thus, the $(n_{\mathrm{RX}},n_{\mathrm{RY}})$th row of $\bm{Z}_{k}$ will be set as $[\frac{l_{\mathrm{R,ave}}^{n_{\mathrm{RX}},n_{\mathrm{RY}}}}{l_{\mathrm{max}}},\frac{w_{\mathrm{R,ave}}^{n_{\mathrm{RX}},n_{\mathrm{RY}}}}{w_{\mathrm{max}}},\frac{h_{\mathrm{R,ave}}^{n_{\mathrm{RX}},n_{\mathrm{RY}}}}{h_{\mathrm{max}}},-1]$, while all the other rows of $\bm{Z}_k$ is set as the zero vector.

\begin{algorithm}
  \caption{The Algorithm to Decode $\hat{\bm{O}}^{\mathrm{B}}_{k}$}
  \label{alg1}
  \begin{algorithmic}[1]
  \REQUIRE ~~\\
  $\hat{\bm{O}}^{\mathrm{B}}_{k}$;\\
  \ENSURE ~~\\
  The optimal BS indices $\bm{b}^{*}$ from VRAN for all the $U$ users;
  \STATE Set $\bm{\mathcal{V}}=\emptyset$ and $\bm{\mathcal{U}}^{'}=\emptyset$;
  \STATE Set $\tilde{i}^{n_{\mathrm{RX}},n_{\mathrm{RY}}}=i^{n_{\mathrm{RX}},n_{\mathrm{RY}}}$, $n_{\mathrm{RX}}=1,2,\cdots,N_{\mathrm{RX}}$, $n_{\mathrm{RY}}=1,2,\cdots,N_{\mathrm{RY}}$;
  \REPEAT
  \STATE Obtain the index of the maximum element in $\hat{\bm{O}}^{\mathrm{B}}_{k}$ as $(n_{\mathrm{RX}}^{\mathrm{max}},n_{\mathrm{RY}}^{\mathrm{max}},b^{\mathrm{max}})$ except the elements with indices in $\bm{\mathcal{V}}$;
  \IF{$\tilde{i}^{n_{\mathrm{RX}}^{\mathrm{max}},n_{\mathrm{RY}}^{\mathrm{max}}}>0$}
  \STATE Obtain a user's BBox inside the $(n_{\mathrm{RX}}^{\mathrm{max}},n_{\mathrm{RY}}^{\mathrm{max}})$th USDF grid as $\mathrm{Box}_{k}^{u^{'}}$ and ensure $u^{'}\notin \bm{\mathcal{U}}^{'}$;
  \STATE Set $b_{u^{'}}^{*}=b^{\mathrm{max}}$;
  \STATE Set $\tilde{i}^{n_{\mathrm{RX}}^{\mathrm{max}},n_{\mathrm{RY}}^{\mathrm{max}}}=\tilde{i}^{n_{\mathrm{RX}}^{\mathrm{max}},n_{\mathrm{RY}}^{\mathrm{max}}}-1$;
  \STATE Set $\bm{\mathcal{U}}^{'}=\bm{\mathcal{U}}^{'}\cup\{u^{'}\}$;
  \STATE Set $\bm{\mathcal{V}}=\bm{\mathcal{V}}\cup \{(n_{\mathrm{RX}},n_{\mathrm{RY}},b^{\mathrm{max}})\ |\ n_{\mathrm{RX}}=1,2,\cdots,N_{\mathrm{RX}}, n_{\mathrm{RY}}=1,2,\cdots,N_{\mathrm{RY}}\}$;
  \ELSE
  \STATE Set $\bm{\mathcal{V}}=\bm{\mathcal{V}}\cup \{(n_{\mathrm{RX}}^{\mathrm{max}},n_{\mathrm{RY}}^{\mathrm{max}},b^{\mathrm{max}})\}$;
  \ENDIF
  \UNTIL{$\mathrm{Card}(\bm{\mathcal{U}}^{'})=U$}.
  \end{algorithmic}
\end{algorithm}

Next, we design the representation form of the output of VRAN. We still adopt the grid form to represent $\bm{b}^{*}$ and $\bm{P}^{*}$, which is consistent with the  previous design of USDF. Specifically, we utilize the $N_{\mathrm{RX}}\times N_{\mathrm{RY}}\times B$ tensor $\bm{O}^{\mathrm{B}}_{k}$ and the $N_{\mathrm{RX}}\times N_{\mathrm{RY}}$ tensor $\bm{O}^{\mathrm{P}}_{k}$ to represent $\bm{b}^{*}$ and $\bm{P}^{*}$, respectively. The $(n_{\mathrm{RX}},n_{\mathrm{RY}})$th row of $\bm{O}^{\mathrm{B}}_{k}$ indicates the optimal indices of BSs for the users inside the $(n_{\mathrm{RX}},n_{\mathrm{RY}})$th USDF grid. If the $(n_{\mathrm{RX}},n_{\mathrm{RY}})$th USDF grid contains the BBoxes of users, then we set $\bm{O}^{\mathrm{B}}_{k}[n_{\mathrm{RX}},n_{\mathrm{RY}},b^{*}_u]=1$, $\forall u\in\bm{\mathcal{U}}_{k}^{n_{\mathrm{RX}},n_{\mathrm{RY}}}$. All the other rows of $\bm{O}^{\mathrm{B}}_{k}$ is set to be the zero vector. The $(n_{\mathrm{RX}},n_{\mathrm{RY}})$th element of $\bm{O}^{\mathrm{P}}_{k}$ indicates the average normalized optimal transmission power for the users inside the $(n_{\mathrm{RX}},n_{\mathrm{RY}})$th USDF grid. For the $(n_{\mathrm{RX}},n_{\mathrm{RY}})$th USDF grid, if the grid contains the BBoxes of users, then we set $\bm{O}^{\mathrm{P}}_{k}[n_{\mathrm{RX}},n_{\mathrm{RY}}]=\frac{1}{i^{n_{\mathrm{RX}},n_{\mathrm{RY}}}}\sum_{u\in\bm{\mathcal{U}}_{k}^{n_{\mathrm{RX}},n_{\mathrm{RY}}}}\frac{P^{*}_u}{P_{\mathrm{max},b^*_u}}$. All the other elements of $\bm{O}^{\mathrm{P}}_{k}$ are set as $0$.

We denote the estimate of $\bm{O}^{\mathrm{B}}_{k}$ and $\bm{O}^{\mathrm{P}}_{k}$ from VRAN as $\hat{\bm{O}}^{\mathrm{B}}_{k}$ and $\hat{\bm{O}}^{\mathrm{P}}_{k}$, respectively. According to the indices of USDF grids that contain users, $\bm{b}^{*}$ can be decoded from $\hat{\bm{O}}^{\mathrm{B}}_{k}$. The detailed steps to decode $\hat{\bm{O}}^{\mathrm{B}}_{k}$ is shown in Algorithm 2. Moreover, $\bm{P}^{*}$ can be determined by the elements of $\hat{\bm{O}}^{\mathrm{P}}_{k}$. Algorithm 2 can avoid the repetition of BS indices for different users, since one BS can only serve at most one user on the same frequency band. It is worth mentioning that when a USDF grid contains the BBoxes of multiple users, the corresponding element in $\bm{O}^{\mathrm{P}}_{k}$ will be used as the transmission power for all the BSs serving these users.

\begin{figure}[t]
\centering
\includegraphics[width=0.91\textwidth]{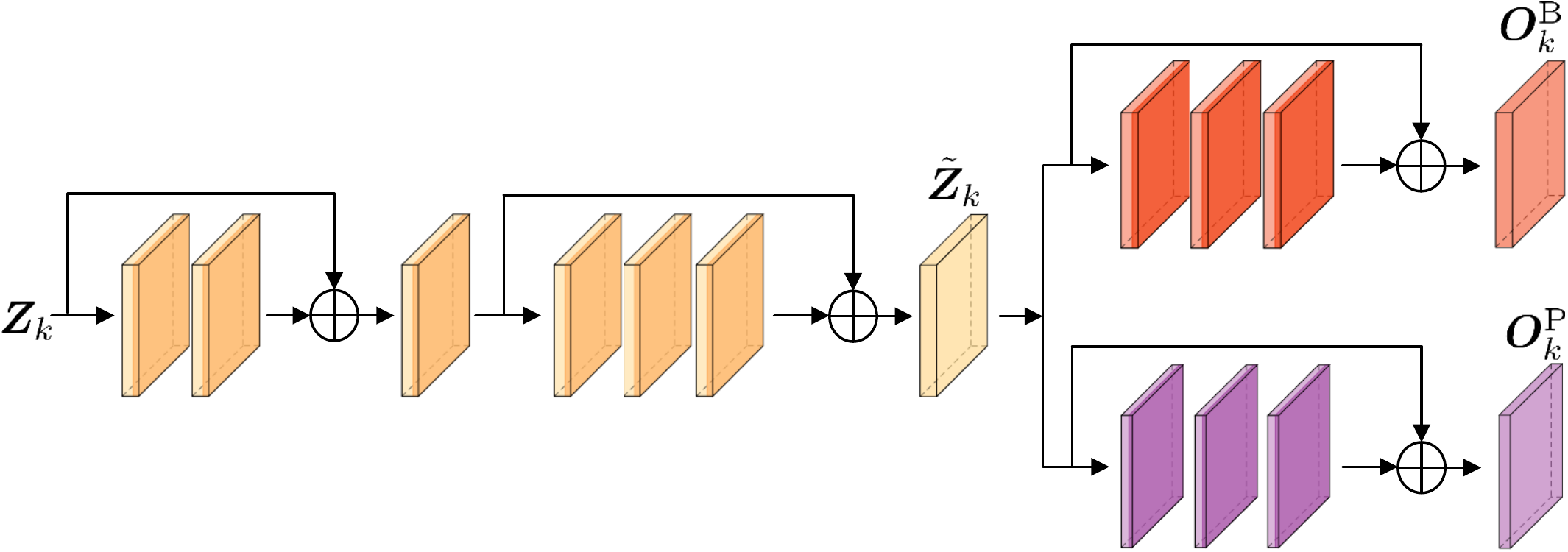}
\caption{The network structure for the proposed VRAN for the mapping from the USDF $\bm{Z}_k$ to both the optimal BS indices $\bm{O}^{\mathrm{B}}_{k}$ and the optimal power allocation $\bm{O}^{\mathrm{P}}_{k}$.}
\end{figure}

To learn the mapping from $\bm{Z}_k$ to both $\bm{O}^{\mathrm{B}}_{k}$ and $\bm{O}^{\mathrm{P}}_{k}$, the network structure of VRAN is designed by the 2D convolutional layers, as shown in Fig.~13. We input $\bm{Z}_k$ into several convolutional layers with the residual connection to obtain a intermediate tensor $\tilde{\bm{Z}}_k$. Then, $\tilde{\bm{Z}}_k$ is fed into two subnetworks with 2D convolutional layers to output $\bm{O}^{\mathrm{B}}_{k}$ and $\bm{O}^{\mathrm{P}}_{k}$, respectively. For each subnetwork, the last convolutional layer is with the Sigmoid activation function.

\begin{table}[t]
\centering
\caption{Vehicle Sizes For Simulation}
\begin{tabular}{|c|c|c|c|}
\hline
\textbf{Type}& \textbf{Length/m}& \textbf{Width/m}& \textbf{Height/m}\\
\hline
Car& 3.71& 1.79& 1.55\\
\hline
Sedan& 4.86& 2.03& 1.65\\
\hline
Van & 5.20& 2.61& 2.47\\
\hline
Bus & 11.08& 3.25& 3.33\\
\hline
\end{tabular}
\end{table}

\section{Simulation Results}
\subsection{Simulation Setup}
To evaluate the performance of the proposed 3DUMM and VBRAM, we construct the communications environment and the moving vehicles to gather the images and channels simultaneously, from which the datasets of UMAN and VRAN can be generated. The details are elaborated as follows:

\subsubsection{Image Generation}
The CARLA\footnote{An open-source autonomous driving simulation platform.} software \cite{carla} is used to produce a nearly real environment model with a crossroad and to generate vehicles, as shown in Fig.~14(a). Moreover, the CARLA can support sensory data collection from multiple different sensors, such as GNSS, camera, and LIDAR. To simulate the traffic flow, we randomly place 15 vehicles at the left and the right traffic lane respectively as the vehicle initialization. Then, all vehicles' motion parameters, including speeds and orientations, are automatically controlled by SUMO\footnote{A traffic simulation software.} \cite{sumo} to generate the traffic flow. We here adopt four vehicles types, including car, sedan, van, and bus, and the corresponding sizes are listed in Table~II.
We put $B=4$ BSs in the four corners of the crossroad respectively, as shown in Fig.~14(b). The $b$th BS is denoted as $\mathrm{BS}_b$. The camera of each BS is installed on the top of the corresponding BS, and all the cameras take images synchronously to monitor the crossroad.

\begin{figure}
  \centering
\subfigure[]{
\begin{minipage}[t]{0.6\linewidth}
\centering
\includegraphics[height=57mm,width=82mm]{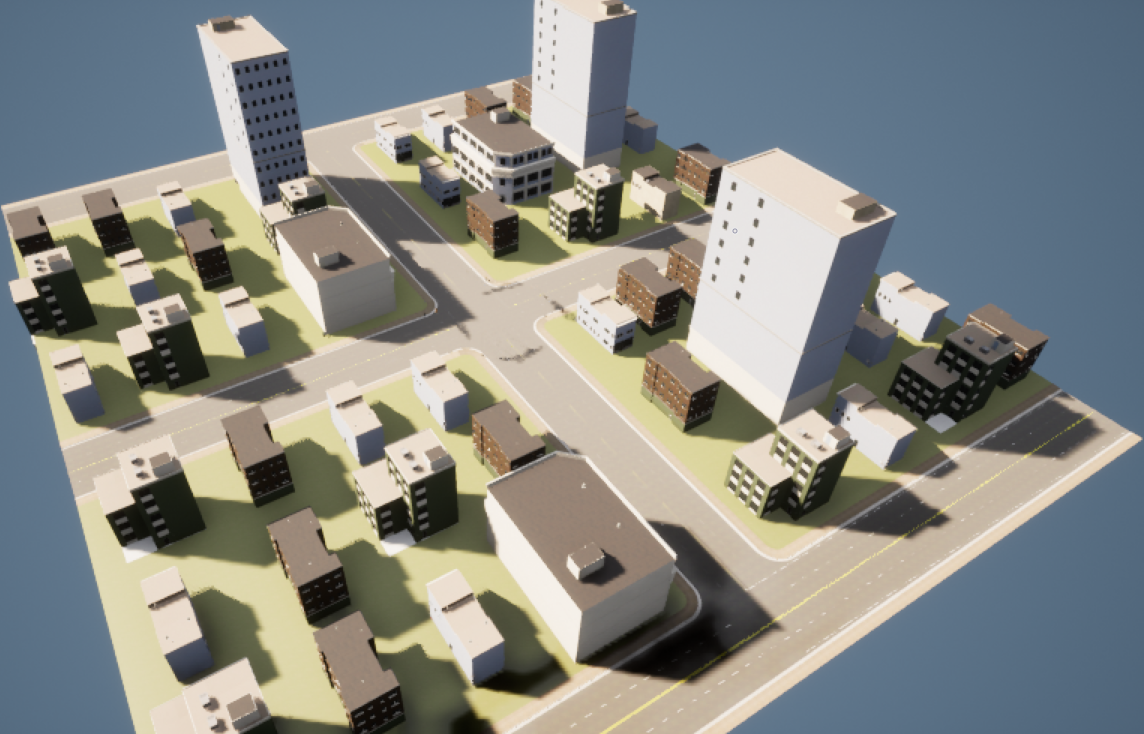}
\end{minipage}%
}%
\subfigure[]{
\begin{minipage}[t]{0.4\linewidth}
\centering
\includegraphics[width=63mm]{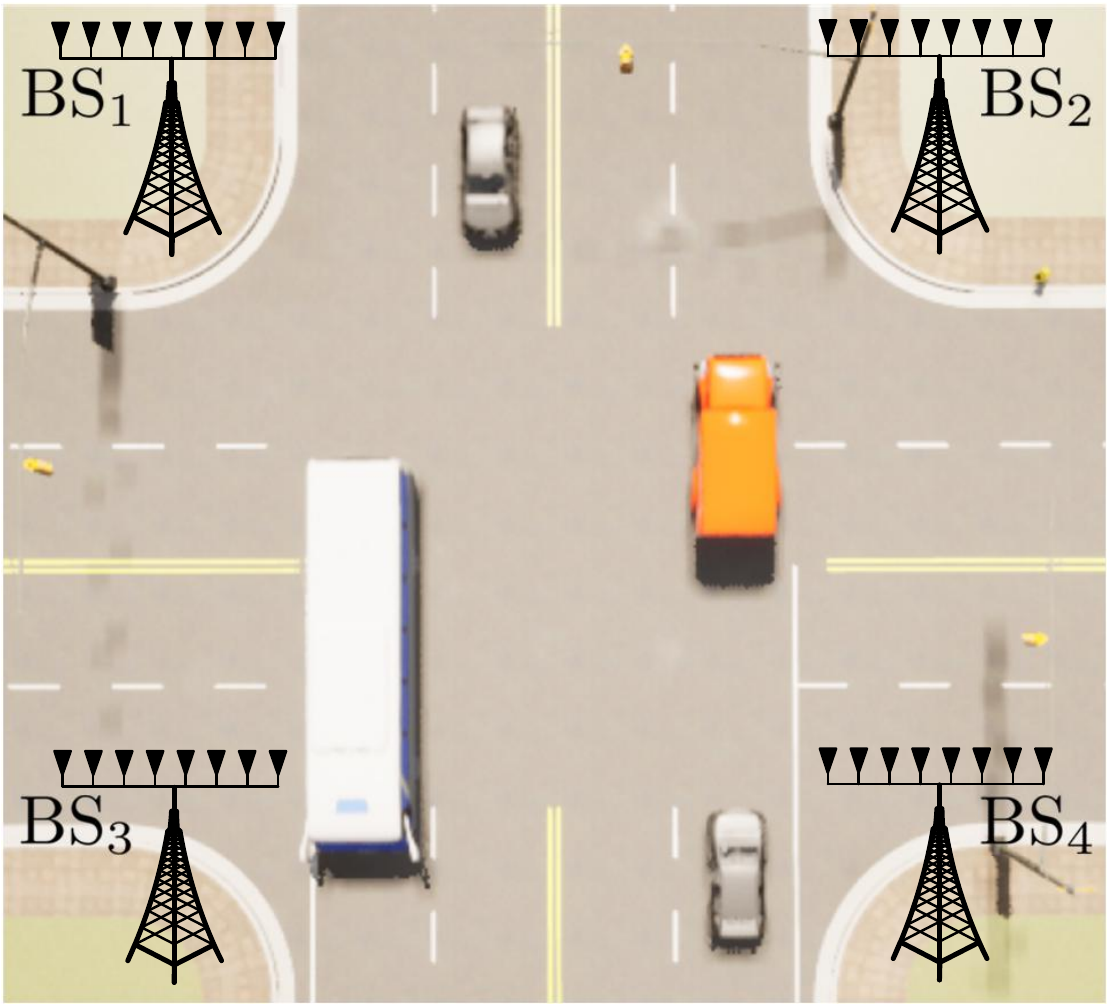}
\end{minipage}
}
\caption{(a) The simulated communications environment in CARLA. (b) The crossroad with 4 BSs.}
\end{figure}
\subsubsection{Channel Generation}
\begin{figure}[t]
\centering
\includegraphics[width=0.6\textwidth]{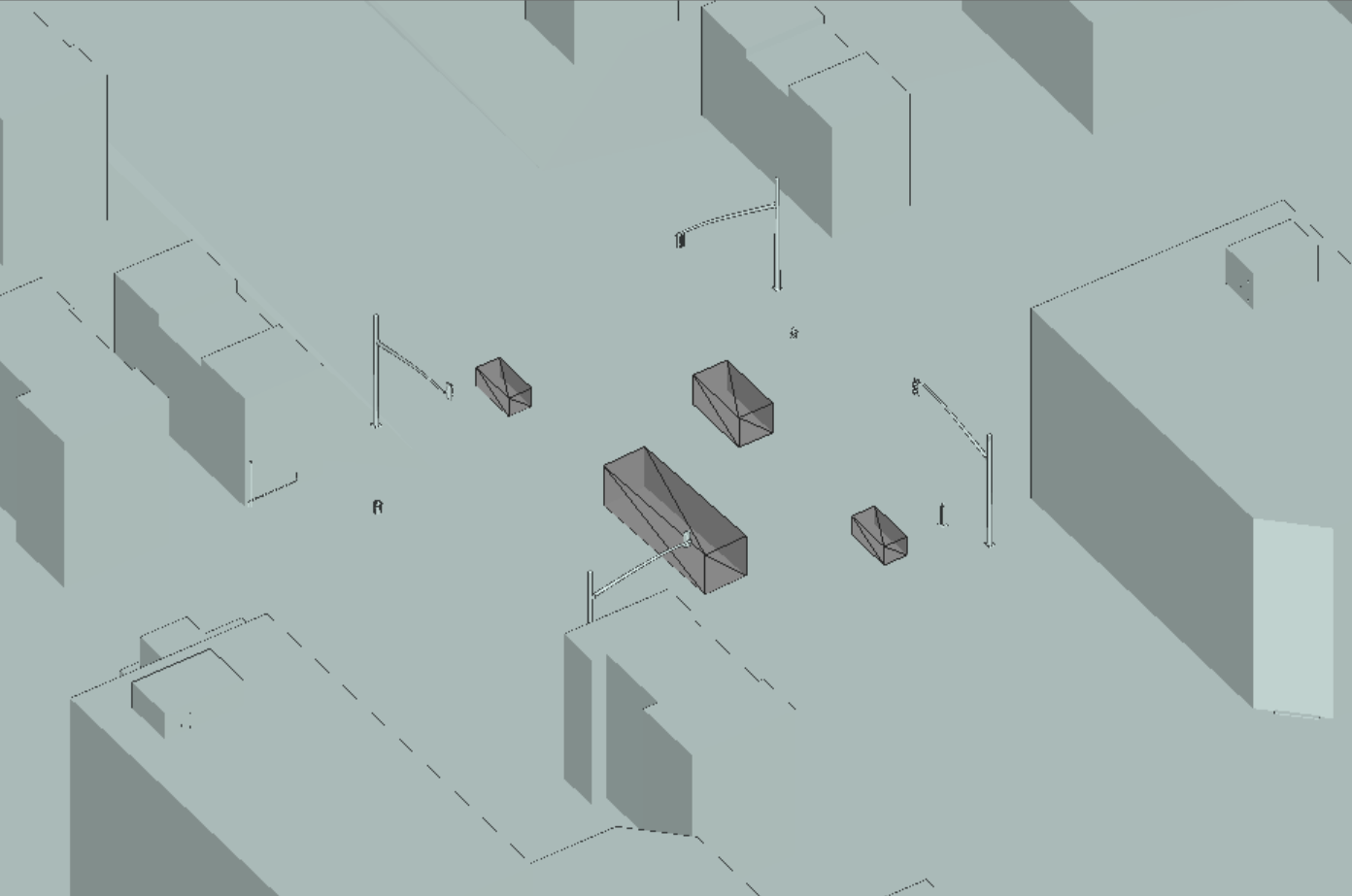}
\caption{The simulated environment in WI. The environmental buildings and vehicles in WI is synchronized with that in CARLA.}
\end{figure}

\subsubsection{Channel Generation}

The classic geometric channel model \cite{AMSa} is adopted to generate the channel matrix
\begin{equation}
\bm{H}_{b,u}=\sum_{l=1}^{L}\alpha_{l,b,u}\bm{a}_\mathrm{r}(\phi^r_{l,b,u})\bm{a}_\mathrm{t}^{\mathrm{H}}(\phi^t_{l,b,u}),
\end{equation}
where $\alpha_{l,b,u}$ is the complex gain of the $l$th path, $\phi^\mathrm{r}_{l,b,u}$ and $\phi^\mathrm{t}_{l,b,u}$ are the $l$th path's azimuth angles of arrival and departure respectively, $\bm{a}_\mathrm{r}(\phi) \in \mathbb{C}^{N_\mathrm{U}\times 1}$ and $\bm{a}_\mathrm{t}(\phi) \in \mathbb{C}^{N_\mathrm{B}\times 1}$ are the complex steering vectors of the receive and transmit ULA array respectively. The antenna spacing of each BS and user is set to be the half carrier wavelength, i.e.,
\begin{equation}
\begin{aligned}
\bm{a}_\mathrm{t}(\phi)&=\frac{1}{\sqrt{N_{\mathrm{B}}}}[1,e^{j\pi \sin(\phi)},\cdots,e^{j(N_{\mathrm{B}}-1)\pi \sin(\phi)}]^{\mathrm{T}},\\
\bm{a}_\mathrm{r}(\phi)&=\frac{1}{\sqrt{N_{\mathrm{U}}}}[1,e^{j\pi \sin(\phi)},\cdots,e^{j(N_{\mathrm{U}}-1)\pi \sin(\phi)}]^{\mathrm{T}}.
\end{aligned}
\end{equation}

The Wireless Insite\footnote{A ray-tracing modeling and simulation software.} (WI) \cite{remcom} is utilized to generate the attenuation/angle/delay parameters of channel paths to obtain the channels that have high spatial correlation with the constructed environment. The environmental buildings and vehicles in WI are completely synchronized with those in CARLA, as shown in Fig.~15. The height of the ULA equipped at each BS is set to be $4.5\mathrm{m}$, and the ULA of each vehicle is set at $0.05\mathrm{m}$ above the roof center of the vehicle. The ULAs of all BSs are set to be parallel with the traffic lane, and the ULA of each vehicle is set to be parallel with the vehicle orientation. For each moment that the $B$ cameras take images, we use WI to produce the channels between all BSs and all vehicles in the crossroad according to the simulation parameters in Table~III.

\begin{table}[t]
\centering
\caption{Critical Parameters of Wireless Insite for Ray Tracing}
\begin{tabular}{|c|c|c|c|}
\hline
\textbf{Parameter}& \textbf{Value}\\
\hline
Carrier Frequency& 28 $\mathrm{GHz}$\\
\hline
Propagation Model& X3D\\
\hline
Building Material& Concrete\\
\hline
Vehicle Material& Metal\\
\hline
Maximum Number of Reflections& 6\\
\hline
Maximum Number of Diffractions& 1\\
\hline
Maximum Paths Per Receiver Point& 25\\
\hline
\end{tabular}
\end{table}

\subsubsection{Dataset Generation}
Since the vehicle trajectories are affected by the vehicle initialization, different vehicle initializations will lead to different spatio-temporal distribution of vehicles. Thus, we carry out $C$ different vehicle initializations to gather a number of different image and channel data to form the datasets of UMAN and VRAN. For the $c$th vehicle initialization, we collect an image set sequence $\bm{\mathcal{I}}_{c,1},\bm{\mathcal{I}}_{c,2},\cdots,\bm{\mathcal{I}}_{c,S}$ of $S$ consecutive shooting moments, where $\bm{\mathcal{I}}_{c,s}$, $c=1,2,\cdots,C$, $s=1,2,\cdots,S$, is an image set containing $B$ images taken by all the cameras at a moment. We denote the moment when cameras taking $\bm{\mathcal{I}}_{c,s}$ as $t_{c,s}$ and denote the number of vehicles in the crossroad at the moment $t_{c,s}$ as $Q_{c,s}$. The channels between the $B$ BSs and the $Q_{c,s}$ vehicles can be obtained according to the channel generation approach in Section V.A-2). Thus, we denote the channel between the $b$th BS and the $q$th vehicle in $Q_{c,s}$ vehicles as $\bm{H}_{c,s,b,q}$. The index $I_{c,s,b,q}$ of the optimal beam pair corresponding to $\bm{H}_{c,s,b,q}$ in the beam pair set $\bm{\mathcal{P}}$ can be calculated by equation (2). Then, the BDF $\bm{D}_{c,s}$ can also be obtained by the image set $\bm{\mathcal{I}}_{c,s}$. Note that we here adopt the \emph{single-stage monocular 3D object detection via keypoint estimation} (SMOKE) \cite{Zliu} as the 3D detection method.

To generate the training/validation/test samples of UMAN, we select the vehicle that satisfies the following two conditions as the user. {\it Condition 1}: the vehicle should exist in the crossroad during the $M$ consecutive moments $t_{c^{'},s^{'}-(M-1)\alpha}$, $t_{c^{'},s^{'}-(M-1)\alpha}$, $\cdots$, $t_{c^{'},s^{'}}$; {\it Condition 2}: $Q_{c^{'},s^{'}}>1$ should hold to avoid the simple scenario with only one vehicle in environment. Next, we obtain the keypoint heatmap $\bm{F}_{c^{'},s^{'},}$ of the selected vehicle by extracting the de-redundancy BBox set $\bm{\mathcal{X}}_{\mathrm{E},c^{'},s^{'}}$ from $\bm{\mathcal{I}}_{c^{'},s^{'}}$. In $\bm{\mathcal{X}}_{\mathrm{E},c^{'},s^{'}}$, the BBox whose plane location is the closest to the location of the selected vehicle is set as the ground truth of the BBox. Then, we denote the indices of the selected vehicle in the $Q_{c^{'},s^{'}-(M-1)\alpha}$, $Q_{c^{'},s^{'}-(M-2)\alpha}$, $\cdots$, $Q_{c^{'},s^{'}}$ vehicles as $q_{c^{'},s^{'},1}$, $q_{c^{'},s^{'},2}$, $\cdots$, $q_{c^{'},s^{'},\alpha}$, respectively.  The beam pair sequence $I_{c^{'},s^{'}-(M-1)\alpha,b,q_{c^{'},s^{'},1}}$, $I_{c^{'},s^{'}-(M-2)\alpha,b,q_{c^{'},s^{'},2}}$, $\cdots$, $I_{c^{'},s^{'},b,q_{c^{'},s^{'},\alpha}}$ and the BDF sequence $\bm{D}_{c^{'},s^{'}-(M-1)\alpha}$, $\bm{D}_{c^{'}, s^{'}-(M-1)\alpha}$, $\cdots$, $\bm{D}_{c^{'},s^{'}}$ can be integrated with $\bm{F}_{c^{'},s^{'},}$ to construct a sample of UMAN for the $b$th BS. All the vehicles that satisfy the above two conditions can be used to construct the samples of UMAN for each BS. We randomly divide the total $C$ vehicle initializations into $C_{\mathrm{train}}$, $C_{\mathrm{valid}}$, and $C_{\mathrm{test}}$ vehicle initializations, where $C=C_{\mathrm{train}}+C_{\mathrm{valid}}+C_{\mathrm{test}}$. The image and channel data collected from the $C_{\mathrm{train}}$, $C_{\mathrm{valid}}$, and $C_{\mathrm{test}}$ vehicle initializations are utilized to generate the training, the validation, and the test datasets of UMAN, respectively.

On the other side, to generate the samples of VRAN, the optimal solution $\bm{b}^*$ and $\bm{P}^*$ of P1 should be obtained. However, since the problem P1 is non-convex, it is hard to obtain the optimal solution\cite{PZhou}. Thus, we here adopt a beam training based resource allocation method (BTRAM) to obtain the sub-optimal solution as the sample label to train VRAN. The BTRAM utilizes the reference signal receiving power (RSRP) from beam training to determine the user scheduling scheme, and then the optimized power allocation is obtained by the traditional WMMSE method \cite{HSun}. Specifically, we firstly set $P_1,P_2,\cdots,P_U$ as the max power $P_{\mathrm{max},b_1},P_{\mathrm{max},b_2},\cdots,P_{\mathrm{max},P_{b_U}}$ for initialization, and then optimize the serving BS indices $\bm{b}$ of all users. We estimate the RSRP $|(\bm{\mathrm{f}}^{\mathrm{U}})^{\mathrm{H}}\bm{H}_{b,u}\bm{\mathrm{f}}^{\mathrm{B}}|^2$, $u\in\bm{\mathcal{U}}$, $b\in\bm{\mathcal{B}}$, $\bm{\mathrm{f}}^{\mathrm{B}}\in \bm{\mathcal{F}}_{\mathrm{B}}$, $\bm{\mathrm{f}}^{\mathrm{U}}\in \bm{\mathcal{F}}_{\mathrm{U}}$, for all beam pairs between any BSs and any users by the beamfroming strategies \cite{MGiordani}. Thus, we can get the RSRP values $\mathrm{RSRP}(b,u,b^{'},u^{'})=|(\bm{\mathrm{f}}_{b^{'},u^{'}}^{\mathrm{U},\mathrm{opt}})^{\mathrm{H}}\bm{H}_{b,u^{'}}\bm{\mathrm{f}}_{b,u}^{\mathrm{B},\mathrm{opt}}|^2$, $b,b^{'}\in\bm{\mathcal{B}}$, $u,u^{'}\in\bm{\mathcal{U}}$, where $\mathrm{RSRP}(b,u,b^{'},u^{'})$ is the receive/interfering power at the $u^{'}$th user from the $b$th BS when the $b$th BS is serving the $u$th user and the $b^{'}$th BS is serving the $u^{'}$th user. Then, the indices of serving BSs for all $U$ users can be determined by the exhaustive search of all $\frac{B!}{(B-U)!}$ permutations of BSs. Once $\bm{b}$ is optimized, the transmission power $\bm{P}$ of all serving BSs can be optimized by the WMMSE method. The steps to solve P1 is shown in Algorithm 3.

\begin{algorithm}
  \caption{The Algorithm to Solve P1}
  \label{alg1}
  \begin{algorithmic}[1]
  \REQUIRE ~~\\
  $\mathrm{RSRP}(b,u,b^{'},u^{'})$, $b,b^{'}\in\bm{\mathcal{B}}$, $u,u^{'}\in\bm{\mathcal{U}}$;\\
  \ENSURE ~~\\
  $\bm{b}^{*}$ and $\bm{P}^{*}$;
  \STATE Obtain all $\frac{B!}{(B-U)!}$ permutations of BS indices by selecting $U$ BSs from the total $B$ BSs, and use these permutations to form a set $\tilde{\bm{\mathcal{P}}}$;\\
  \STATE Obtain the optimal BS permutation $\bm{b}^{*}$ by
  \begin{equation*}
  \bm{b}^{*}= \mathop{\arg\max}_{\bm{b}\in \tilde{\bm{\mathcal{P}}}}\sum_{u\in\bm{\mathcal{U}} }\log_2(1+\frac{P_{\mathrm{max},b_{u}}\mathrm{RSRP}(b_u,u,b_u,u)}{\sum_{u^{'}\in \bm{\mathcal{U}}\backslash\{u\}}P_{\mathrm{max},b_{u^{'}}}\mathrm{RSRP}(b_{u^{'}},u^{'},b_u,u)+\sigma^2}).
  \end{equation*}
  \STATE Replace $\bm{b}$ with $\bm{b}^{*}$ in (11), optimize $\bm{P}$ by the WMMSE method and obtain $\bm{P}^{*}$.
  \end{algorithmic}
\end{algorithm}

Then, to generate the training and validation samples of VRAN, we select the moment $t_{\tilde{c}^{'},\tilde{s}^{'}}$ that can meet the requirement $Q_{\tilde{c}^{'},\tilde{s}^{'}}\geq U$. The $U$ users are selected from the $Q_{\tilde{c}^{'},\tilde{s}^{'}}$ vehicles, and the $\frac{Q_{\tilde{c}^{'},\tilde{s}^{'}}!}{(Q_{\tilde{c}^{'},\tilde{s}^{'}}-U)!U!}$ different combinations of users can be obtained. For each combination, we assume the BBox is accurately matched and obtain the corresponding USDF by the image set $\bm{\mathcal{I}}_{\tilde{c}^{'},\tilde{s}^{'}}$. The corresponding solution of P1 can be calculated by the simulated user channels and BTRAM. The obtained USDF and the solution of P1 are used to construct a sample of VRAN. Thus, for each the moment when the number of vehicles is greater than or equal to $U$, all the user combinations are obtained to construct the samples for producing the training and validation dataset of VRANs. To generate the test samples of VRAN, the BBox of each user is obtained by the proposed 3DUMM. For the moment $t_{\tilde{c}^{'},\tilde{s}^{'}}$, we count the number of consecutive moments when $Q_{\tilde{c}^{'},\tilde{s}^{'}}$ vehicles exist in the crossroad, and thereby determine the input sequence length $M$ of UMAN to match each vehicle with the corresponding BBox. For each vehicle, the matching process is conducted by a randomly selected BS. Then, the USDF for the test sample of VRAN can be obtained by the matched user BBoxes and the other scatters' BBoxes. The test dataset of VRAN is still generated by extracting different user combinations from $Q_{\tilde{c}^{'},\tilde{s}^{'}}$ vehicles, which is consistent with the generation approach of the training and the validation set of VRAN. We use $C_{\mathrm{train}}$, $C_{\mathrm{valid}}$, and $C_{\mathrm{test}}$ vehicle initializations that are consistent with UMAN to generate the training, the validation, and the test datasets of VRAN, respectively.

\subsection{Performance Metric}
The user matching accuracy (UMAC) is adopted as the metric to evaluate the performance of 3DUMM. We denote the test dataset of UMAN as $\bm{\mathcal{T}}_{\mathrm{UMAN}}$ and denote the number of test samples in $\bm{\mathcal{T}}_{\mathrm{UMAN}}$ with the BBox accurately matched as $\tau_{\mathrm{AC}}$. Then, we obtain $\mathrm{UMAC}=\frac{\tau_{\mathrm{AC}}}{\mathrm{Card}(\bm{\mathcal{T}}_{\mathrm{UMAN}})}$.

We adopt the achievable transmission rate ratio (ATRR) as the metric to analyze the accuracy of VBRAM.  We denote the test dataset of VRAN as $\bm{\mathcal{T}}_{\mathrm{VRAN}}$. Moreover, for the $\tau$th test sample in $\bm{\mathcal{T}}_{\mathrm{VRAN}}$, we denote the total transmission rate achieved by VBRAM as $\bm{R}^{\mathrm{VRAN}}_{\tau}$ and denote the total transmission rate achieved by the BTRAM as $\bm{R}^{\mathrm{opt}}_{\tau}$. Thus, we obtain $\mathrm{ATRR}=\frac{\sum_{\tau=1}^{\mathrm{Card}(\bm{\mathcal{T}}_{\mathrm{VRAN}})}\bm{R}^{\mathrm{VRAN}}_{\tau} }{\sum_{\tau=1}^{\mathrm{Card}(\bm{\mathcal{T}}_{\mathrm{VRAN}})}\bm{R}^{\mathrm{opt}}_{\tau}}$.

\begin{figure}[t]
\centering
\includegraphics[width=0.7\textwidth]{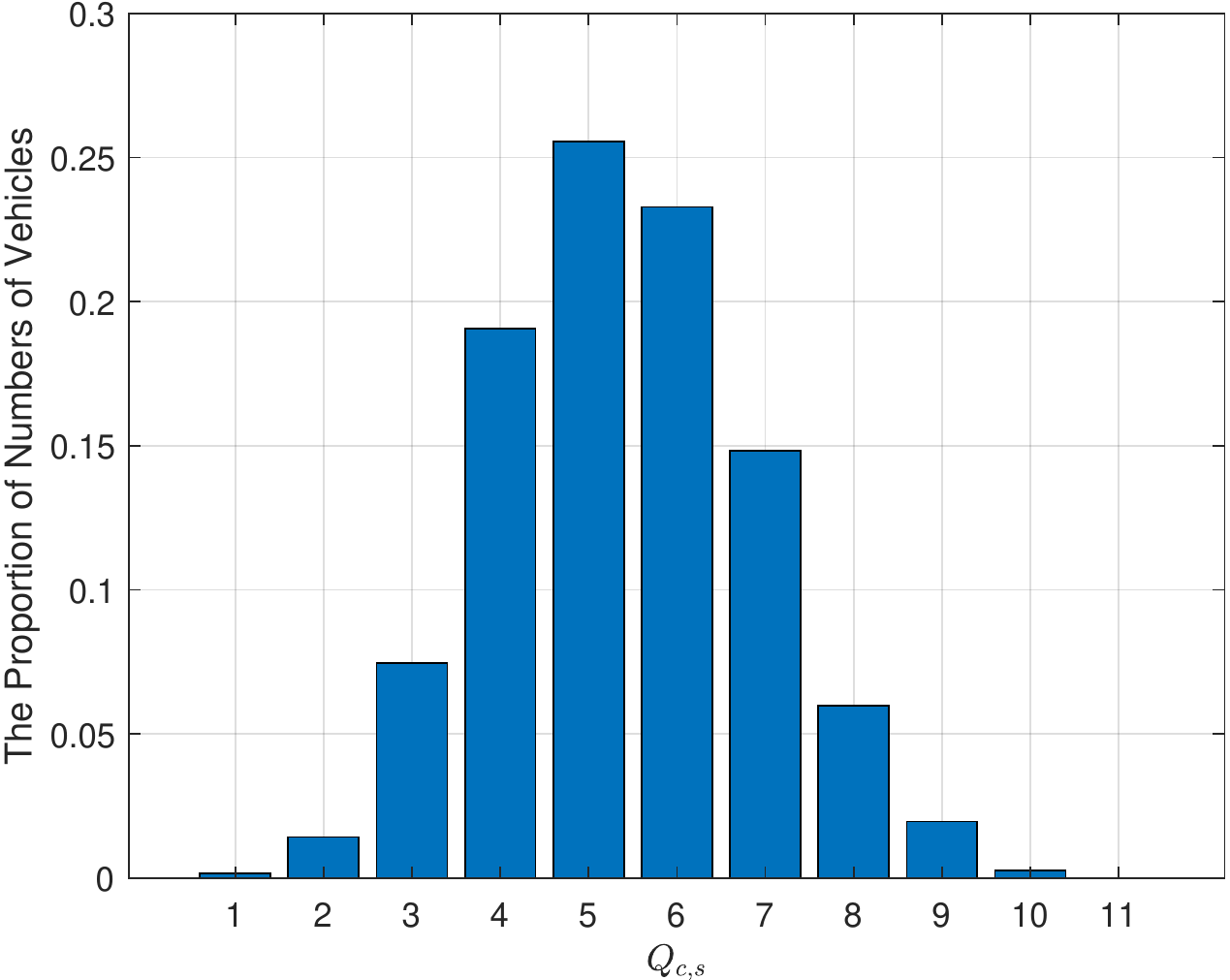}
\caption{The proportions of different numbers of vehicles in simulation.}
\end{figure}
\subsection{Simulation Parameters}
The parameters settings are: $N_\mathrm{B}=64$ and $N_\mathrm{U}=16$, respectively; $N_{\mathrm{CB}}=N_\mathrm{B}=64$; $N_{\mathrm{CU}}=N_\mathrm{U}=16$; $\bm{\mathrm{f}}^{\mathrm{B},i}=\bm{a}_{\mathrm{t}}(\frac{2i-2-N_{\mathrm{CB}}}{2N_{\mathrm{CB}}}\pi)$, $i=1,2,\cdots,N_{\mathrm{CB}}$ and $\bm{\mathrm{f}}^{\mathrm{U},i}=\bm{a}_{\mathrm{r}}(\frac{2i-2-N_{\mathrm{CU}}}{2N_{\mathrm{CU}}}\pi)$, $i=1,2,\cdots,N_{\mathrm{CU}}$; $\gamma=\frac{1}{3}$; $T_{\mathrm{f}}=0.05\mathrm{s}$; $\alpha=5$; BCT length $T_{\mathrm{b}}$ is set to be fixed and $T_{\mathrm{b}}=\alpha T_{\mathrm{f}}$; $L_{\mathrm{D}}$ and $W_{\mathrm{D}}$ are set to be $0.52 \mathrm{m}$ and $0.44 \mathrm{m}$ respectively; $ N_{\mathrm{DX}}=40$ and $N_{\mathrm{DY}}=160$; $L_{\mathrm{F}}=2L_{\mathrm{D}}$ and $W_{\mathrm{F}}=2W_{\mathrm{D}}$; $N_{\mathrm{FX}}=20$ and $N_{\mathrm{FY}}=80$; $\tilde{\gamma}=0.3$; $L_{\mathrm{R}}=L_{\mathrm{D}}$ and $W_{\mathrm{R}}=W_{\mathrm{D}}$; $ N_{\mathrm{RX}}=20$ and $N_{\mathrm{RY}}=80$;
$\frac{P_{\mathrm{max},b}}{\sigma^2\sum_{c=1}^{C}\sum_{s=1}^{S}Q_{c,s}}\sum_{c=1}^{C}\sum_{s=1}^{S}\sum_{q=1}^{Q_{c,s}}||\bm{H}_{c,s,b,q}||_F^2=25\mathrm{dB}, b\in\bm{\mathcal{B}}$; $C$ is set as 50; $C_{\mathrm{train}}$, $C_{\mathrm{valid}}$, and $C_{\mathrm{test}}$ are set to be 40, 5, and 5 respectively; $S$ is set as 300. According to the numbers $Q_{c,s}$, $c=1,2,\cdots,C$, $s=1,2,\cdots,S$, of vehicles at different moments, the proportions of different numbers of vehicles are counted and shown in Fig.~16. It is seen that three to eight vehicles exist in the environment most of the time.

For UMAN, three convolutional layers with the kernel size (3,3) and the stride (1,1) are utilized to obtain the intermediate tensor $\tilde{\bm{D}}_k$ from the stacked BDF sequence $\check{\bm{D}}_{k}$. The number of filters of the three convolutional layers are 16, 16, and 32 respectively, and the input of the second convolutional layer is added to the output of the third convolutional layer by the residual connection. An embedding layer with output dimension 1024 and an LSTM layer with output dimension 2048 are utilized to obtain the intermediate tensor $\tilde{\bm{v}}_k$ from the input beam sequence $I_{k-M+1},I_{k-M+2},\cdots,I_{k}$. Three convolutional layers with the kernel size (3,3) and the stride (1,1) is utilized to obtain the intermediate tensor $\tilde{\bm{I}}_k$ from $\tilde{\bm{v}}_k$. For the three convolutional layers, the number of filters are 8, 16, and 32 respectively, and the input of the second convolutional layer is added to the output of the third convolutional layer by the residual connection. Then, an average pooling layer with pool size (2,2) and three convolutional layers with the kernel size (3,3) and the stride (1,1) are used to obtain the heatmap $\bm{F}_k$ from the sum of $\tilde{\bm{D}}_k$ and $\tilde{\bm{I}}_k$. For the three convolutional layers, the number of filters are 16, 8, and 1 respectively, and the input of the first convolutional layer is added to the output of the second convolutional layer by the residual connection. The batch normalization and ReLU activation function are used for all the above convolutional layers except the last convolutional layer with Sigmoid activation function. The UMAN is trained for 60 epochs, and the loss function is set as the penalty-reduced focal loss \cite{XZhou}. Since the points around a keypoint can be regarded as good estimation of the keypoint, the penalty-reduced focal loss decreases the loss penalty of the points around the ground-truth keypoints to reduce the learning difficulty of DNN. Specifically, the penalty-reduced focal loss of the designed UMAN for heatmap estimate $\hat{\bm{F}}_k$ is given by
\begin{equation}
L_{\mathrm{UMAN}}=-\sum_{n_{\mathrm{FX}}}^{N_{\mathrm{FX}}}\sum_{n_{\mathrm{FY}}}^{N_{\mathrm{FY}}}\left\{
\begin{aligned}
&(1-\hat{\bm{F}}_k[n_{\mathrm{FX}},n_{\mathrm{FY}}])^{\beta}\log(\hat{\bm{F}}_k[n_{\mathrm{FX}},n_{\mathrm{FY}}]),\ \bm{F}_k[n_{\mathrm{FX}},n_{\mathrm{FY}}]=1,\\
&(1-\bm{F}_k[n_{\mathrm{FX}},n_{\mathrm{FY}}])^{\eta}(\hat{\bm{F}}_k[n_{\mathrm{FX}},n_{\mathrm{FY}}])^{\beta}\log(1-\hat{\bm{F}}_k[n_{\mathrm{FX}},n_{\mathrm{FY}}]),\ \mathrm{otherwise},
\end{aligned}\right.
\end{equation}
where $\beta$ and $\eta$ are the adjustable parameters, and the term $(1-\bm{F}_k[n_{\mathrm{FX}},n_{\mathrm{FY}}])^{\eta}$ is used to reduce the loss penalty for the points around the ground-truth keypoints. We set $\beta$ and $\eta$ as 2 and 4 respectively.

For VRAN, 17 convolutional layers with the kernel size (3,3) and the stride (1,1) are used to obtain the intermediate tensor $\tilde{\bm{Z}}_k$ from the USDF $\bm{Z}_k$. For the 15 convolutional layers, the number of filters are 16, 16, 16, 16, 32, 32, 64, 64, 128, 128, 256, 256, 512, 512, 1024, 1024, and 2048 respectively. The inputs of the 2nd, 4th, 6th, 8th, 10th, 12th, 14th, and 16th convolutional layer are added to the outputs of the 3rd, 5th, 7th, 9th, 11th, 13th, 15th, and 17th convolutional layer respectively by the residual connection. Then, a subnetwork composed by 13 convolutional layers with the kernel size (3,3) and the stride (1,1) is used to obtain the BS indices tensor $\bm{O}_k^{\mathrm{B}}$ from $\tilde{\bm{Z}}_k$. The number of filters of the 13 convolutional layers are 1024, 512, 512, 256, 256, 128, 128, 64, 64, 32, 32, 16, and 4 respectively. The inputs of the 1st, 3rd, 5th, 7th, 9th, and 11th convolutional layer are added to the outputs of the 2nd, 4th, 6th, 8th, 10th, and 12th convolutional layer respectively by the residual connection. A subnetwork composed by 7 convolutional layers with the kernel size (3,3) and the stride (1,1) is used to obtain the power allocation tensor $\bm{O}_k^{\mathrm{P}}$ from $\tilde{\bm{Z}}_k$. For the 7 convolutional layers, the number of filters are 128, 64, 64, 32, 32, 16, and 1 respectively. The inputs of the 1st, 3rd, and 5th convolutional layer are added to the outputs of the 2nd, 4th, and 6th convolutional layer respectively by the residual connection. The batch normalization and ReLU activation function are used for all the convolutional layers except for the last convolutional layers with Sigmoid activation function of the two subnetworks. The VRAN is trained for 20 epochs. The form of output $\bm{O}_k^{\mathrm{B}}$ of VRAN is similar to the heatmap form of $\bm{F}_k$, as both $\bm{O}_k^{\mathrm{B}}$ and $\bm{F}_k$ are designed by grid division. Hence, the loss function for the estimate $\hat{\bm{O}}_k^{\mathrm{B}}$ is designed according to the penalty-reduced focal loss of equation (16). Moreover, note that only the elements corresponding to the USDF grids with users are required to decode $\hat{\bm{O}}_k^{\mathrm{B}}$. Thus, the loss function for $\hat{\bm{O}}_k^{\mathrm{B}}$ is given by
\begin{equation}
\begin{aligned}
&L_{\mathrm{VRAN},\mathrm{B}}=-\frac{1}{B\sum_{n_{\mathrm{RX}}}^{N_{\mathrm{RX}}}\sum_{n_{\mathrm{RY}}}^{N_{\mathrm{RY}}}\mathbbm{1}(i^{n_{\mathrm{RX}},n_{\mathrm{RY}}}>0)}\\
&\cdot\sum_{n_{\mathrm{RX}}}^{N_{\mathrm{RX}}}\sum_{n_{\mathrm{RY}}}^{N_{\mathrm{RY}}}\sum_{b=1}^{B}\left\{
\begin{aligned}
&(1-\hat{\bm{O}}_k^{\mathrm{B}}[n_{\mathrm{RX}},n_{\mathrm{RY}},b])^{\beta}\log(\hat{\bm{O}}_k^{\mathrm{B}}[n_{\mathrm{RX}},n_{\mathrm{RY}},b]),\ \bm{O}_k^{\mathrm{B}}[n_{\mathrm{RX}},n_{\mathrm{RY}},b]=1,\ i^{n_{\mathrm{RX}},n_{\mathrm{RY}}}>0,\\
&-(\hat{\bm{O}}_k^{\mathrm{B}}[n_{\mathrm{RX}},n_{\mathrm{RY}},b])^{\beta},\ \bm{O}_k^{\mathrm{B}}[n_{\mathrm{RX}},n_{\mathrm{RY}},b]\neq 1,\ i^{n_{\mathrm{RX}},n_{\mathrm{RY}}}>0,\\
&0,\ \mathrm{otherwise},
\end{aligned}\right.
\end{aligned}
\end{equation}
where the indicator function $\mathbbm{1}(\mathcal{C})$ is 1 when the condition $\mathcal{C}$ is true, or is 0 otherwise.
The loss function for the estimate $\hat{\bm{O}}_k^{\mathrm{P}}$ is given by
\begin{equation}
L_{\mathrm{VRAN},\mathrm{P}}=\frac{1}{\sum_{n_{\mathrm{RX}}}^{N_{\mathrm{RX}}}\sum_{n_{\mathrm{RY}}}^{N_{\mathrm{RY}}}\mathbbm{1}(i^{n_{\mathrm{RX}},n_{\mathrm{RY}}}>0)}\sum_{n_{\mathrm{RX}}}^{N_{\mathrm{RX}}}\sum_{n_{\mathrm{RY}}}^{N_{\mathrm{RY}}}\left\{
\begin{aligned}
&(\bm{O}_k^{\mathrm{P}}[n_{\mathrm{RX}},n_{\mathrm{RY}}]-\hat{\bm{O}}_k^{\mathrm{P}}[n_{\mathrm{RX}},n_{\mathrm{RY}}])^{\beta},\ i^{n_{\mathrm{RX}},n_{\mathrm{RY}}}>0,\\
&0,\ \mathrm{otherwise}.
\end{aligned}\right.
\end{equation}

\subsection{Results and Discussions}
\begin{table}[t]
\centering
\caption{List of all the proposed and compared methods}
\begin{tabular}{|c|c|c|c|}
\hline
\textbf{Abbreviation}& \textbf{Method}\\
\hline
3DUMM& 3D detection based user matching method\\
\hline
MCUMM& Multi-class classification based user matching method\\
\hline
RUMM& Random user matching method\\
\hline
VBRAM& Vision based resource allocation method\\
\hline
BTRAM& Beam training based resource allocation method\\
\hline
NBBRAM& Nearest BS based resource allocation method\\
\hline
RRAM& Random resource allocation method\\
\hline
\end{tabular}
\end{table}

We first analyze the learning performance of UMAN. The losses of UMAN for $\mathrm{BS}_1$ on validation set versus the number of training epochs are shown in Fig.~17 under different input sequence length $M$ of UMAN. It is seen that with the increasing of the number of training epochs, the losses of UMAN under different $M$ all reduce and reach convergence. Moreover, the smaller convergent loss of UMAN can be achieved under the larger $M$. This is because with the increasing of $M$, the more visual information and the beam information of user is provided to support the more accurate user matching. For the computation complexity, the number of weight parameters and FLOPs of UMAN depends on the input sequence length $M$. For $M=1$, the number of weight parameters of UMAN is approximately $3.94\times 10^7$ and the number of FLOPs of UMAN is approximately $2.27\times 10^8$. The number of weight parameters increases by 432 and the number of FLOPs increases by about $5.5\times 10^6$ when $M$ increases by 1. It can be seen that the FLOPs for $M=5$ increases by $9.7\%$ compared with that for $M=1$, while the UMAC for $M=5$ increases by $4.2\%$ compared with that for $M=1$ on average. This indicates there is a tradeoff between matching accuracy and computation overhead. Furthermore, to perform the proposed DNNs in real time, the professional GPU, such as the NVIDIA RTX A6000 or RTX 4090, and the application-specific integrated circuit (ASIC) chip \cite{ASIC} designed for the neural network are usable in practice. For instance, we adopt the NVIDIA RTX 3090 to implement the DNNs, and the inference time for one sample of UMAN is approximately $43.8\mathrm{ms}$ on average for different $M$. Moreover, some neural network structure optimization methods, such as neural network pruning \cite{SLin} and knowledge distillation \cite{GHinton}, can be used to reduce the computational complexity and improve the inference speed of the neural network.

\begin{figure}[t]
	\begin{minipage}[t]{0.5\linewidth}
		\centering
	\includegraphics[width=83mm]{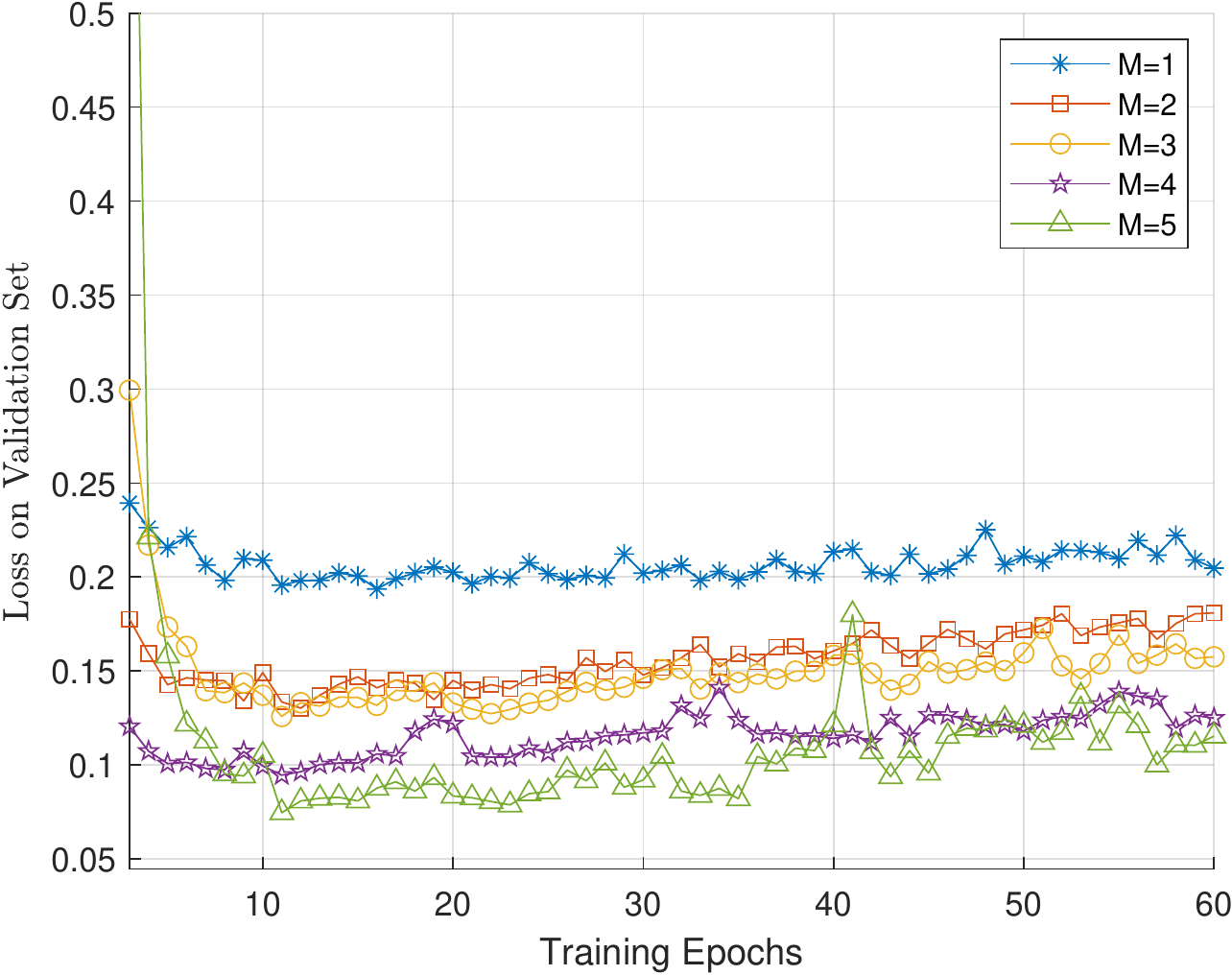}
		\caption{The losses of UMAN for the $\mathrm{BS}_1$ on validation set under different $M$ versus the increase of the number of training epochs.}
	\end{minipage}
	\hspace{1ex}
	\begin{minipage}[t]{0.5\linewidth}
		\centering
			\includegraphics[width=83mm]{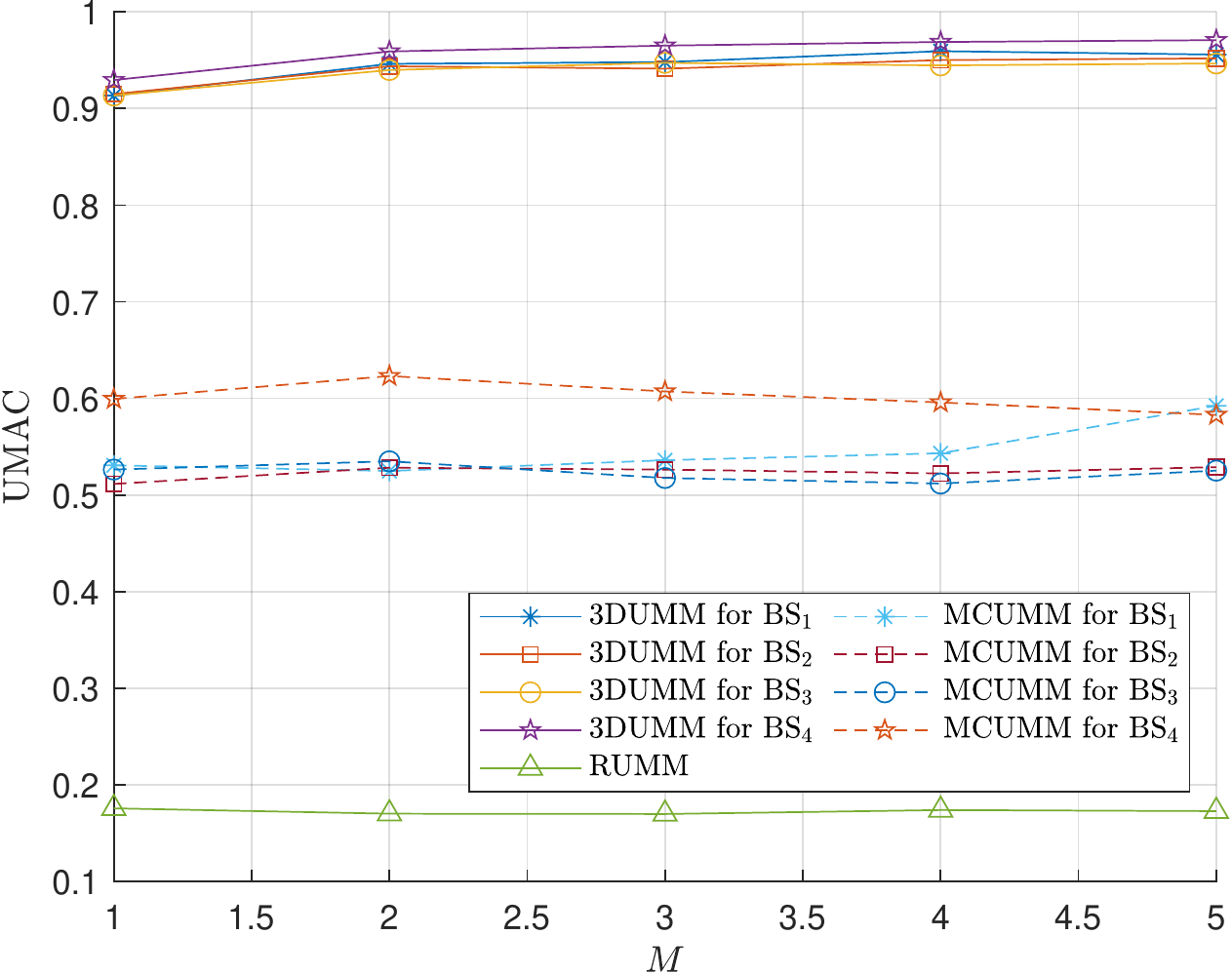}
		\caption{The $\mathrm{UMAC}$ of the 3DUMM, MCUMM and RUMM for all the 4 BSs under different $M$.}
	\end{minipage}
\end{figure}

We further evaluate the user matching performance of UMAN. The best epoch with the minimal loss over the validation set is selected, and we utilize the weight corresponding to the best epoch as the optimal weight of UMAN. We compare the proposed 3DUMM with the multi-class classification based user matching method (MCUMM) in \cite{VMPinho} and the random user matching method (RUMM). The MCUMM fixes the number of vehicles in the environment and utilizes the one-hot encoding to represent the index of the BBox of user. Hence, for the sake of fairness, the final three convolutional layers of UMAN are replaced by their fully connected layers with node numbers 1000, 400, and $O_{\mathrm{max}}$ respectively, where $O_{\mathrm{max}}=12$ is the maximum value of $\mathrm{Card}(\bm{\mathcal{X}}_{\mathrm{E},c^{'},s^{'}})$ and $\bm{\mathcal{X}}_{\mathrm{E},c^{'},s^{'}}$ is the de-redundancy BBox set used for dataset generation. According to the coordinates at the $\mathrm{Y}_{\mathrm{G}}$-axis, the BBoxes in $\bm{\mathcal{X}}_{\mathrm{E},c^{'},s^{'}}$ are sorted in decreasing order for indexing. Next, RUMM will randomly select a BBox in $\bm{\mathcal{X}}_{\mathrm{E},c^{'},s^{'}}$ as the BBox of the user. For convenience of reference, the abbreviations of all the proposed and compared methods are listed in the Table~IV.

\begin{figure}[t]
\centering
\includegraphics[width=0.7\textwidth]{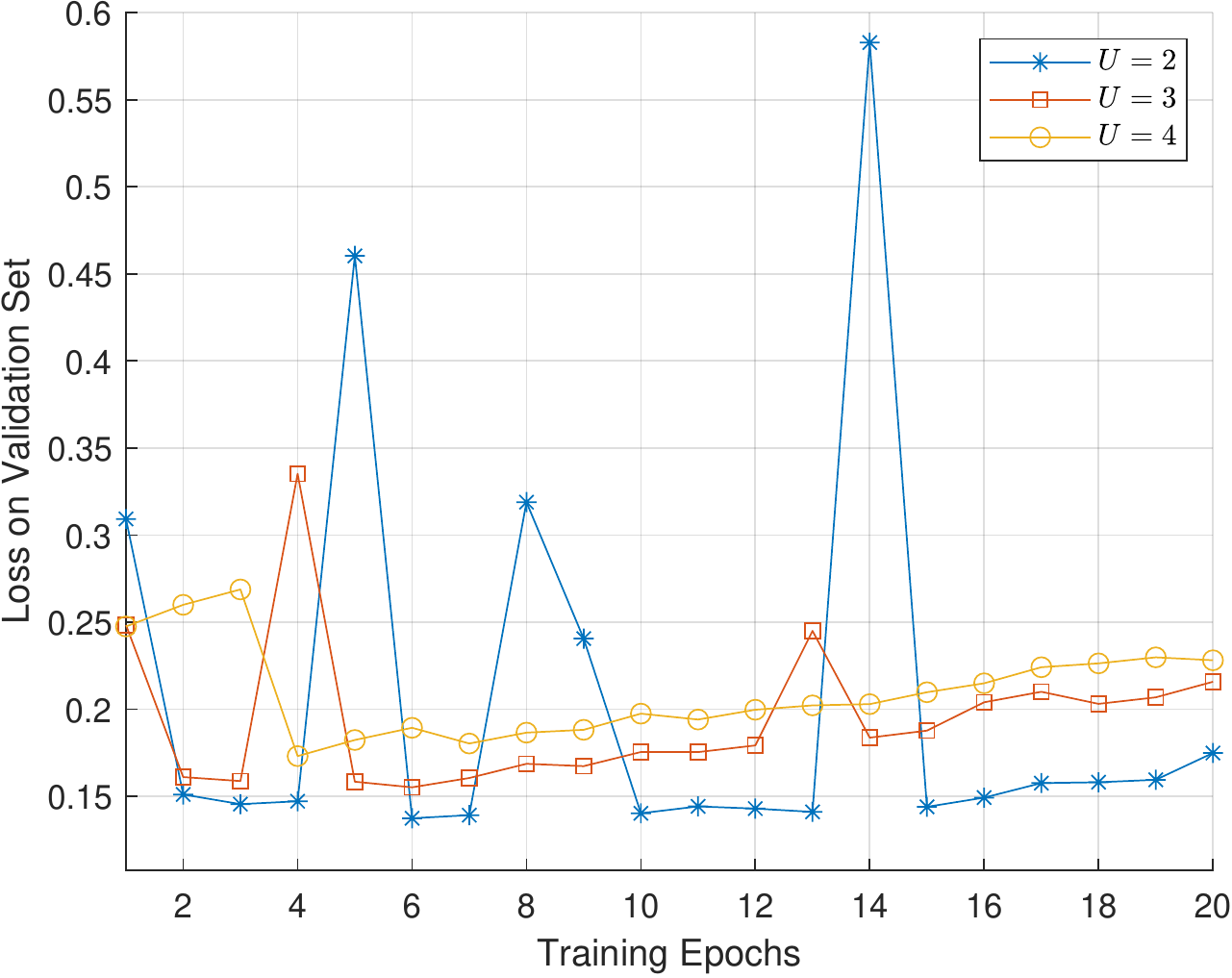}
\caption{The losses of VRAN for different $U$ on validation set versus the increase of the number of training epochs.}
\end{figure}

\begin{figure}[t]
	\begin{minipage}[t]{0.5\linewidth}
		\centering
	\includegraphics[width=82mm]{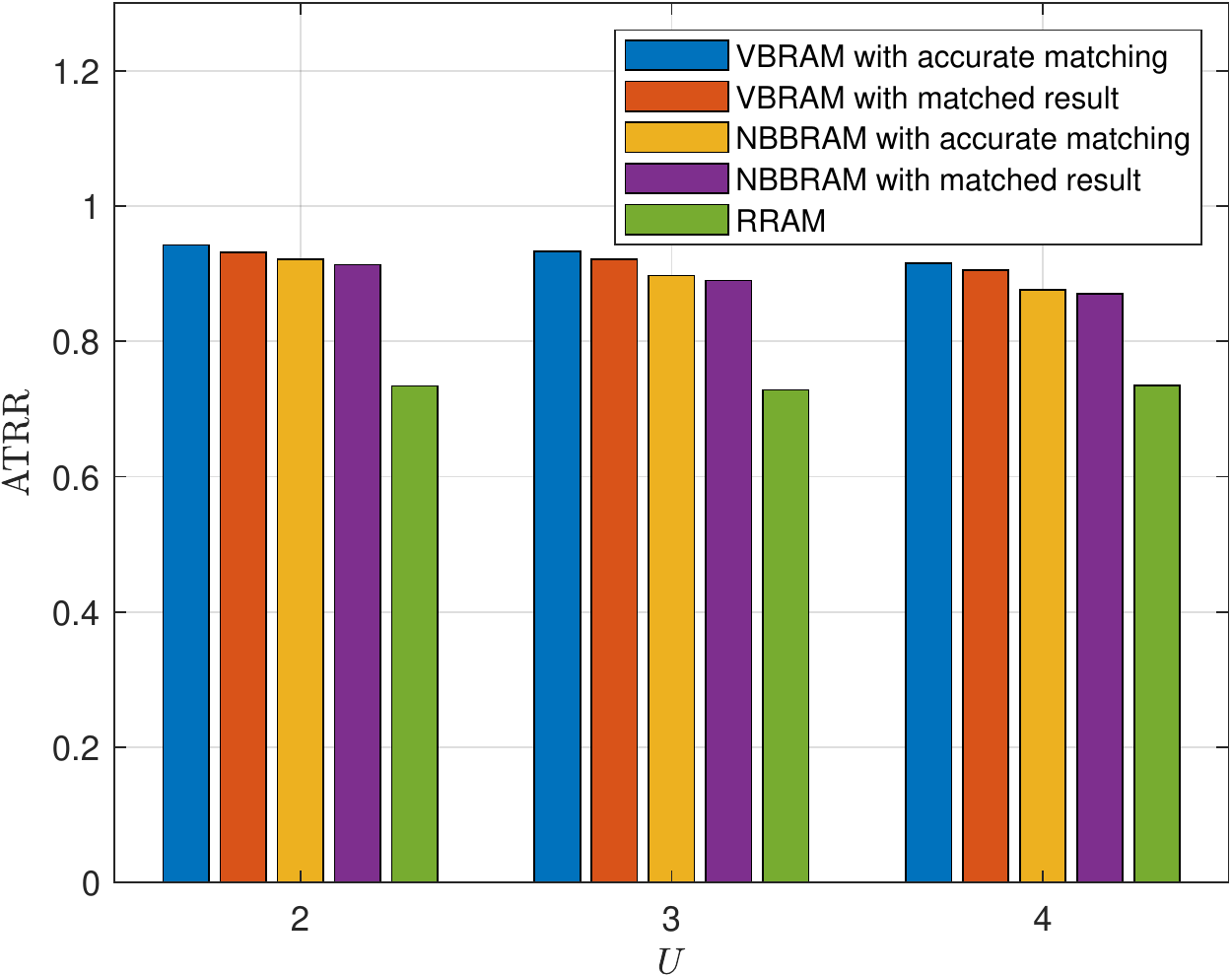}
		\caption{$\mathrm{ATRR}$ of VBRAM, NBBRAM and RRAM for different $U$.}
        \label{VBS_simu2}
	\end{minipage}
	\hspace{1ex}
	\begin{minipage}[t]{0.5\linewidth}
		\centering
			\includegraphics[width=83mm]{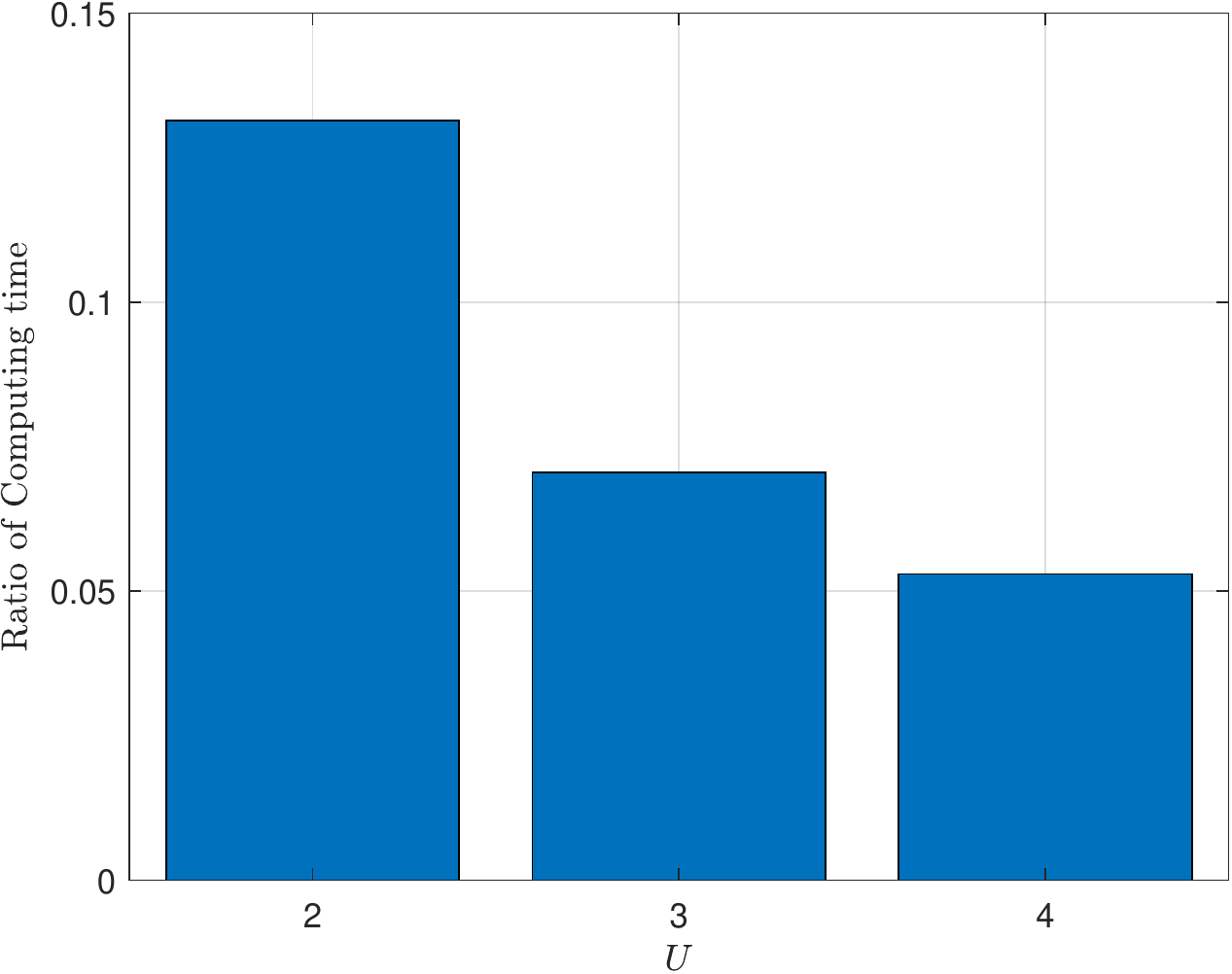}
		\caption{Ratios of computing time between VBRAM and BTRAM under different $U$.}
    \label{VBS_simu3}
	\end{minipage}
\end{figure}

The matching accuracy $\mathrm{UMAC}$ of 3DUMM, MCUMM, and RUMM for all the 4 BSs under different $M$ are shown in Fig.~18. It can be seen that the $\mathrm{UMAC}$ of 3DUMM is approximately $40\%$ and $75\%$ higher than that of MCUMM and RUMM, respectively. This indicates that the heatmap containing user location distribution is significantly superior to the one-hot encoding for user matching when the number of vehicles is varying. It is also seen that with the increasing of $M$, the $\mathrm{UMAC}$ of 3DUMM for all 4 BSs can increase, since the larger $M$ can provide richer information of the location distribution and beam of user, while the $\mathrm{UMAC}$ of the 4 BSs are slightly different. The highest $\mathrm{UMAC}$ $97.1\%$ is achieved by the 3DUMM for $\mathrm{BS}_4$, which implies that the proposed 3DUMM can well handle the user matching problem under the environment with multiple vehicles.

Next, we analyze the learning performance of VRAN. The losses of VRAN for different $U$ on validation set versus the number of training epochs are shown in Fig.~19. It is seen that VRAN for different $U$ is trained to reach convergence gradually. With the increasing of $U$, the loss of VRAN also increases, since it is more difficult to learn the optimal resource allocation scheme with multiple users. For the computation complexity of VRAN, the number of weight parameters of VRAN is approximately $7.28\times 10^{7}$ and the number of FLOPs of VRAN is approximately $9.31\times 10^{11}$. The inference time for one sample of VRAN is approximately $156.6\mathrm{ms}$ in the simulation.

We compare the proposed VBRAM with the \emph{nearest BS} based resource allocation method (NBBRAM) and the random resource allocation method (RRAM). For each user, NBBRAM will select the BS with the smallest distance from the user as the corresponding serving BS, and the center location of the corresponding BBox matched by 3DUMM is used as the estimation of the user location. If multiple users are closest to the same BS, the BS will only serve the user with the minimum distance from it and every other user is arranged to connect to the second nearest BS. RRAM will randomly select a BS index permutation from $\tilde{\bm{\mathcal{P}}}$ as $\bm{b}$, where $\tilde{\bm{\mathcal{P}}}$ is the set containing all permutations of BS indices. Both NBBRAM and RRAM set the transmission power $P_{b_u}$ as the maximum transmission power $P_{\mathrm{max},b_u}$, $u\in\bm{\mathcal{U}}$.

Then, the transmission rate ratio $\mathrm{ATRR}$ of VBRAM, NBBRAM and RRAM for different $U$ are shown in Fig.~20. Moreover, VBRAM and NBBRAM with accurate user matching are also used for comparison. It can be seen that the $\mathrm{ATRR}$ of VBRAM with matched result can reach approximately $92\%$ on average for different $U$ and is $18.7\%$ higher than that of RRAM. Thus, though the UMAC realized by 3DUMM is approximately $95.2\%$ on average, VBRAM with matched result can achieve comparable transmission rate to that of the traditional BTRAM without the huge cost of time and spectrum, while the BTRAM relies on the RSRP to realize sub-optimal resource allocation. The $\mathrm{ATRR}$ of VBRAM with matched result can be $1.8\%$, $3.1\%$ and $3.5\%$ higher than that of NBBRAM with matched result for $M=2,3$, and $4$ respectively. This indicates that though both VBRAM and NBBRAM rely on the visual information, the neural network of VBRAM is superior to the nearest BS selection strategy of NBBRAM, because NBBRAM does not consider the influence of surrounding scattering objects. The superiority of neural network becomes more evident with the increasing of $U$. It is also seen that the ATRR of VBRAM and NBBRAM with accurate user matching are only $1.1\%$ and $0.7\%$ higher than the ATRR of VBRAM and NBBRAM with matched result, respectively, which indicates that the proposed 3DUMM can support the performance of VBRAM and NBBRAM to approach the same performance achieved by accurate user matching.

To reveal the superiority of the VBRAM in terms of computing time, the ratios of computing time between VBRAM and BTRAM under different $U$ are shown in Fig.~21. It can be seen that only approximately $5.3\%$ computing time of BTRAM is needed by VBRAM to perform the resource allocation for $U=4$. Moreover, the ratio of computing time decreases with the increase of $U$. The reason is that more iterations are necessary to be calculated by BTRAM for more users and thereby lead to more computing time. However, as the dimensions of the input $\bm{Z}_k$ and the outputs $\bm{O}_k^{\mathrm{B}}$ and $\bm{O}_k^{\mathrm{P}}$ of VRAN are independent to $U$, the computation overhead of VRAN is not affected by $U$. These results demonstrate VBRAM can effectively reduce the computational time overhead of resource allocation compared with the traditional method, especially for the scenario with a large number of users.

\section{Conclusion}
We propose a vision aided communication scheme that including a user matching method and a resource allocation method. The proposed 3DUMM can be applicable to the environment with varying number of objects by estimating the user location distribution and significantly outperform the conventional multi-class classification based method. Compared with the traditional resource allocation method, the proposed VBRAM can realize commensurate transmission rate and have lower spectrum and computing time overhead. In fact, the joint utilization of different modal user characteristic information, such as the beam, the RSRP, and the channel state information of the user, can further improve the user matching accuracy and is worthy of research in the future.

\balance

\end{document}